\documentclass[aps,prxquantum, twocolumn,superscriptaddress,groupedaddress]{revtex4}  

\usepackage{stmaryrd}
\usepackage{physics}
\usepackage{graphicx}
\usepackage{color}
\usepackage{bm,bbold}
\usepackage{bbm}
\usepackage{enumerate}
\usepackage{color}
\usepackage{xcolor}
\usepackage{float}
\usepackage{comment}
\usepackage{multirow}
\graphicspath{figures} 
\usepackage{amsfonts, amsmath, amsthm, amssymb}

\usepackage[colorlinks,bookmarks=false,citecolor=blue,linkcolor=red,urlcolor=blue]{hyperref}
\begin{document}
\title{Three-outcome multipartite Bell inequalities: applications to dimension witnessing and spin-nematic squeezing in many-body systems}

\author{Guillem M\"uller-Rigat}
\email{guillem.muller@icfo.eu}
\affiliation{ICFO-Institut de Ciencies Fotoniques, The Barcelona Institute of Science and Technology, Castelldefels (Barcelona) 08860, Spain}

\author{Albert Aloy}
\affiliation{Institute for Quantum Optics and Quantum Information, Austrian Academy of Sciences, Boltzmanngasse 3, A-1090 Vienna, Austria}
\affiliation{Vienna Center for Quantum Science and Technology (VCQ), Faculty of Physics, University of Vienna, Vienna, Austria}

\author{Maciej Lewenstein}
\affiliation{ICFO-Institut de Ciencies Fotoniques, The Barcelona Institute of Science and Technology, Castelldefels (Barcelona) 08860, Spain.}
\affiliation{ICREA, Pg.~Lluís Companys 23, 08010 Barcelona, Spain.}

\author{Matteo Fadel}
\email{fadelm@phys.ethz.ch}
\affiliation{Department of Physics, ETH Z\"{urich}, 8093 Z\"{urich}, Switzerland}

\author{Jordi Tura}
\affiliation{$\langle aQa^L \rangle$ Applied Quantum Algorithms, Universiteit Leiden, Leiden, The Netherlands}
\affiliation{Instituut-Lorentz, Universiteit Leiden, P.O. Box 9506, 2300 RA Leiden, The Netherlands}

\begin{abstract}
\textbf{ABSTRACT.~} We present a three-outcome Bell inequality which we show to be naturally suited to explore nonlocal correlations in many-body spin-1 systems or SU(3) models. In the specific, we demonstrate from this inequality an experimentally practical Bell correlation test based on the measurement of collective spin components. Next, we analyze such an inequality as a dimension witness, namely criteria whose violation signals the impossibility of reproducing the inferred correlations by Hilbert spaces of a certain dimension. In particular, our approach enables the certification of genuine three-level correlations in systems of arbitrary size, which cannot be reproduced by an ensemble of qubits. Then, we demonstrate how metrological resources, in the form of spin-nematic squeezed states as produced in spin-1 Bose Einstein condensates involving thousands of atoms, maximally violate such a witness. Moreover, we propose that the robustness of the violation can be used to certify the number of high-dimensional parties in the system. In this direction, we demonstrate that, in the thermodynamic limit, the squeezed correlations described above require the system to
be composed entirely of qutrits. Our results shed light
on the fundamental role of high-dimensional quantum resources in many-body systems and their application for metrological purposes.

\end{abstract}

\maketitle

\section*{INTRODUCTION}

The formalism of quantum mechanics brought revolutionary ideas that challenged preconceived notions in classical physics, such as locality and realism. A characteristic feature of quantum mechanics is that any complex linear combination of two valid quantum states results in an another valid, qualitatively different, quantum state. Moreover, when considering multipartite quantum systems, such characteristic feature may lead to correlations that violate the principles of locality and realism \cite{EPR35}, resulting in one of the strongest departure from classical physics. In 1964, the seminal work of J. S. Bell offered a precise mean to characterize correlations that follow local-realistic principles by formalizing local hidden variable models and, in turn, the so-called Bell inequalities \cite{Bell1964}. Correlations that violate a Bell inequality are incompatible with local hidden variable theories, and are thus called nonlocal.  

Nowadays, nonlocal correlations constitute one of the building blocks of modern physics \cite{BrunnerRMP2014}. Besides their role in the history of quantum mechanics \cite{EPR35,Bell1964} and their fundamental interest \cite{PawlowskiNature2009, FritzNatComms2013, NavascuesNatComms2015}, these correlations have driven a paradigm shift in quantum protocols \cite{LydersenNatPhot2010} by introducing Device-Independent (DI) Quantum Information Processing tasks \cite{AcinDIQKD}. That is, tasks that do not require trust nor knowledge about their physical implementation. For instance, nonlocality serves as a resource for DI tasks such as self-testing \cite{MayersYaoST2004, SATWAP, WangScience2018, Kaniewski2018}, randomness amplification and expansion \cite{ColbeckRenner2012, Colbeck,PironioNature2010}, DI quantum key distribution \cite{AcinDIQKD, Arnon-FriedmanNatComms2018}, DI detection and certification of entanglement \cite{BancalDIGME, MoroderPRL2013, YCLiangentDepth, BaccariPRX2017, AloyPRL2019} or nonlocality depth \cite{BaccariPRA2019}.

From a more applied perspective, recent progresses in quantum technologies, such as in quantum computing and simulation, have reached control over large-scale quantum systems composed by hundreds of constituent elements \cite{Deutsch2020}. What makes these platforms depart from their classical analogue are genuine quantum resources, e.g. entanglement and nonlocality. It is thus extremely relevant the scalable detection and characterization of quantum correlations in these controllable many-body systems \cite{FrerotROPP2023}. Furthermore, we are in the advent of high-dimensional quantum resources, extending beyond qubits, with promising applications in enhancing the efficiency of certain algorithms~\cite{Ringbauer2022, Chi2022}. In a similar spirit, multimode (or multilevel) many-body systems provide benefits for metrological applications~\cite{Fadel_NJP2023}. For example, the balanced spin-1 Dicke state takes advantage of all three components and thus displays higher interferometric sensitivity~\cite{zou2018beating}. The recent progress in this direction prompts the need for rigorous and experimentally-accessible dimension witnesses, that is, criteria whose violation signals the impossibility of reproducing the inferred correlations by a Hilbert space of a certain dimension.

However, revealing these quantum correlations is notoriously hard, especially in multipartite scenarios.
The underlying mathematical complexity \cite{Pitowski1989} (NP-complete in the general case \cite{BabaiCompCompl1991}) has impeded a full characterization of nonlocality beyond the simplest scenarios \cite{PitowskiPRA2001, SliwaPLA2003, PironioJPA50}. Nevertheless, significant progress has been achieved in this direction for the bipartite scenario~\cite{Collins2004,Oudot2018}. More generally, specific Bell inequalities have been designed for important classes of quantum sates (\textit{e.g.,} graph states \cite{BellIneqsGraphStates, BellIneqsGraphStates2, BASTA}, symmetric states \cite{Bancal2010, Bancal2012}, continuous variable systems \cite{CavalcantiCVBellIneqs, BellIneqsCVMeasurements}, etc.), or Bell Inequalities with desirable properties have been put forward (\textit{e.g.,} achieving maximal violation by maximally entangled \cite{MayersYaoST2004, SATWAP, Kaniewski2018} or partially entangled states \cite{BampsPRA2015, YangPRAR2013, ColadangeloNatComms2017}). In a series of works \cite{SciencePaper,AnnPhys}, a successful approach to derive practical multipartite Bell inequalities (BI) was based on restricting to permutationally invariant (PI) few-body correlators. Such PIBIs allowed for the experimental detection of Bell correlations in many-body quantum systems using collective { measurements} only \cite{SchmiedScience2016, EngelsenPRL2017}.

Here, we propose a three-outcome PIBI resulting from the systematic examination of the set of three-outcome local-realistic correlations in the many-body scenario, which, to date, remains largely unexplored. Already for systems involving only two parties, the first derivation of a Bell inequality involving more than two outcomes \cite{CollinsPRL2002} had fundamental implications in showing that entanglement and nonlocality are not equivalent, since maximally entangled states did not maximally violate the mentioned inequality. However, the derivation of multipartite Bell inequalities involving more than two outcomes remains largely unexplored, especially because of the underlying complexity of the problem. The increased number of outcomes results in a prohibitive scaling of the number of valid Bell inequalities, which makes their enumeration intractable. Therefore, even if some three-outcome multipartite Bell inequalities have been obtained \cite{alsina2016operational}, the state-of-the-art is far from being in a true many-body regime of experimental relevance. 

Three-outcome Bell inequalities are also naturally fitted to explore the role of nonlocal correlations in three-level many-body systems, such as paradigmatic three-level models \cite{LipkinNucPhys1965,GnutzmannJPhysA1999} or spin-$1$ atomic ensembles \cite{HoPRL98,LawPRL98}. For instance, one may consider spin-1 Bose-Einstein condensates routinely implemented in the lab using $^{87}$Rb atoms \cite{barrett2001all,schmaljohann2004dynamics}, where quantum correlations among the particles appears in the form of spin-nematic squeezing Ref.~\cite{HamleyNat12}.

In addition, for Bell inequalities with more than two outcomes, Jordan lemma~\cite{Jordan1875}, which ensures that the maximal quantum violation (Tsirelson bound) is attained with qubits, no longer applies. Therefore, higher dimensional systems, such as qutrits, may be necessary to saturate its maximal quantum expectation value. {This fact prompts us to analyze our Bell inequality violation as a dimension witness, which certifies the minimal dimension of the single-particle Hilbert space, in addition to certifying Bell nonlocality.}

The problem of testing the Hilbert space dimension from the statistics alone~\cite{Gallego2010,Kaszlikowski2002} was first considered in Ref.~\cite{Brunner2008}. Later, Refs.~\cite{navascues2015bounding,navascuesPRA2015} characterized this set via convex optimization techniques by adding suitable dimension constraints to the celebrated Navascués-Pironio-Acín hierarchy \cite{NavascuesNJP2008}. However, the increasing computational complexity of these approaches makes them unfeasible to implement for more than a few parties. For instance, in Ref.~\cite{GrandjeanPRA2012} tripartite dimension witnesses are introduced. Here, we exploit the symmetries of the PIBI to scale Bell dimension witness to the many-body regime. In specific, our approach enables the certification of genuine three-level correlations in systems of arbitrary size, which cannot be reproduced e.g. by an ensemble of qubits. 

In order to be of interest for DI tasks, Bell inequalities need not only to be violated (i.e. detect nonlocality), but it is also desirable to have the possibility of characterizing which states and measurements yield their maximal quantum violation and how to test these in the laboratory. { As an essential asset of this work, we provide practical witnesses based on accessible collective observables. Surprisingly, we find that the maximally violating state displays exquisite sensitivity in the form of spin-nematic squeezing~\cite{HamleyNat12} and it is naturally produced in spin-$1$ atomic ensembles~\cite{HoPRL98,LawPRL98} and exploited for metrological applications~\cite{PezzeRMP2018}. In addition, we show that in the thermodynamic limit these squeezed correlations cannot be reproduced by an ensemble of qubits, contrary to what one might naively expect. \\

\section*{RESULTS}

\subsection*{Permutation-invariant Bell inequalities (PIBIs)}
\label{sec:PIBIs}

We consider a $(n,k,r)$ Bell scenario, i.e $n$ parties, each of them having access to { $k$} measurement settings with $r$ outcomes, see Figure~\ref{fig:Bell_scenario}. After several rounds of collecting results, they are able to infer one- and two-body marginal probability distributions 
\begin{equation}
    \vec{p} = \{p_i(a|x),\   p_{ij}(ab|xy)\} .
    \label{eq:pij}
\end{equation}
Here, $p_i(a|x)$ is the probability for party $i\in [n]$ to obtain outcome $a\in [r]$ when measuring $x\in[k]$. Likewise, $p_{ij}(ab|xy)$ is the conditional probability for parties $(i,j)$ to obtain outcomes $(a,b)$ by measuring $(x, y)$, respectively. In the following, we will refer to these marginal probability distributions or to other statistics that may be inferred from the experiment as \textit{data}. \\

\begin{figure}[h]
\centering
\includegraphics[width=0.9\linewidth]{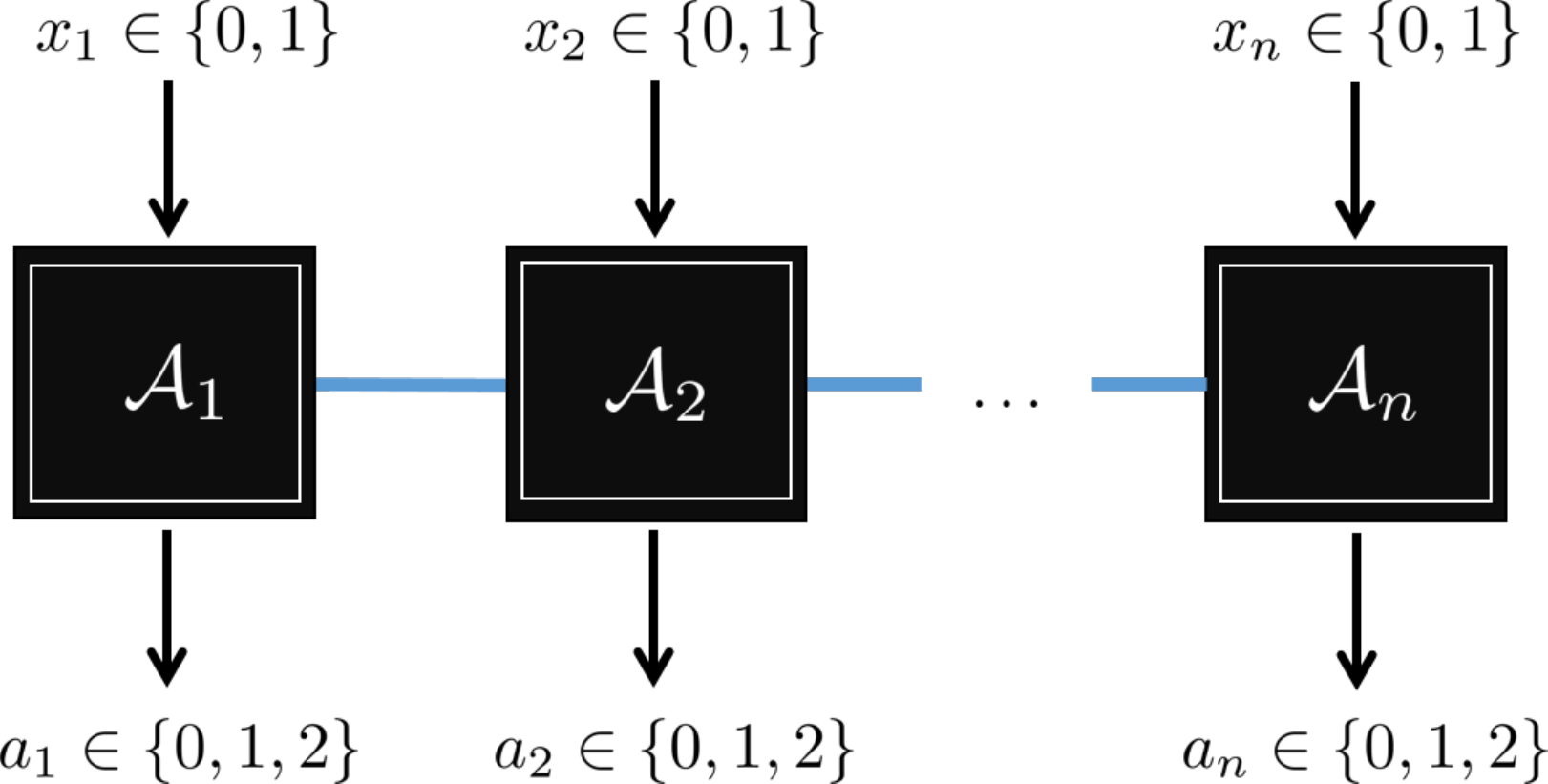}
\caption{Sketch of the $(n,2,3)$ Bell scenario. Each party $\mathcal{A}$ is represented as a \emph{black box}, to highlight ignorance about its inner physical description. In every run of the experiment, each party $i\in [n]$ independently chooses to perform measurement $x_i$ on their share of the system and then register the resulting outcome $a_i$. After enough repetitions, the collected results can be used to infer the joint statistics, e.g. Eq.~\eqref{eq:pij}, from which quantum correlations can be detected.}
\label{fig:Bell_scenario}
\end{figure}

The correlations Eq.~\eqref{eq:pij} might admit a classical explanation, namely trough the so-called local-hidden variable models (LHVMs). In such a case, up to shared randomness, the correlations factorize 
\begin{equation}
    p_{ij}(ab|xy) = \int_{\lambda\in \Lambda} p_i(a|x, \lambda)p_j(b|y, \lambda)d\mu(\lambda)   ,
    \label{eq:probs}
\end{equation}
where $\{d\mu(\lambda)\}_{\lambda\in \Lambda}$ is a probability measure. The set of probabilities $\vec{p}$ may be interpreted as a real vector in an affine space. In such a geometric picture, the set of all LHVM correlations forms a \textit{polytope} $\mathbb{P}$, i.e. a convex set with finite number of vertices. This is delimited by linear inequalities of the form $B = \vec{\alpha}\cdot\vec{p}-\beta_c\geq 0 $, namely \textit{Bell inequalities} (BIs) with coefficients $\vec{\alpha}$ and classical bound $\beta_c$. The minimal set of such inequalities describe the facets of $\mathbb{P}$. As BIs are necessarily fulfilled for any $\vec{p}$ described by LHVMs, the violation of one of them signals \emph{Bell nonlocality} or, in other words, the impossibility to explain the inferred statistics through LHVMs. If the data $\vec{p}$ is derived from a quantum state via the Born rule, the violation of a BI also certifies quantum entanglement.    \\

In this work, we focus on families of BIs that are scalable in the number of parties $n$. To construct some of these, we impose symmetries in the Bell scenario \cite{SciencePaper, GuillemPRXQuantum}. Specifically, we consider probabilities that are averaged over all permutations of the subsystems as 
\begin{equation}
  \label{eq:Pcol}
\begin{split}
    \mathcal{P}_{a|x} &= \sum_{i\in [n]}{p_i(a|x)} \\
    \mathcal{P}_{ab|xy} &=\sum_{i\neq j\in [n]}{p_{ij}(ab|xy)}.
\end{split} 
\end{equation}
Note that we may refer this terms as probabilities, despite the fact that they have a different normalization convention, $\sum_{a\in [3]}\mathcal{P}_{a|x} = n$. These averaged one- and two- body probabilities are routinely measured in current experiments ranging from cold atoms, trapped ions to photonic platforms or superconducting systems~\cite{FriisNatRevPhys2018,FrerotROPP2023}. In contrast, generic higher-order correlations (such as $N$-body probabilities) may be more informative but also much more demanding to infer. In this work, we restrict to the information contained in statistics Eq.~\eqref{eq:Pcol} to build tools to diagnose quantum correlations and high-dimensional quantum resources in such many-body systems. In this framework, the local polytope $\mathbb{P}$, as defined earlier, is projected onto the space of permutation-invariant correlations Eq.~\eqref{eq:Pcol}. This projection is also a polytope, $\mathbb{P}_S$, which contains all the permutation-invariant (PI) correlations admitting a LHVM explanation. The exponential growth in the number of parties $n$ of the dimension of $\mathbb{P}$ deems our considered scenario intractable in the general case. However, the projection operation reduces drastically the combinatorial complexity of the problem and it allows for explicit construction of $\mathbb{P}_S$ for a few parties $n$. \\

To find Bell inequalities that hold for any $n$, a possible approach is to fully solve the local polytope $\mathbb{P}_S$ for small $n$, and infer patterns in the coefficients of the inequalities \cite{SciencePaper,WagnerPRL2017,GuoPRL23}. Our goal is to do this in the $(n,2,3)$ scenario, but the complexity of the problem does not allow us to go further than $n=3$ parties. { After imposing additional symmetries, we fully solve the resulting polytope until $n\sim 17$ with reasonable computational power (see Methods). In doing so, we infer the Bell inequality
\begin{equation}
\begin{split}
    B =& \mathcal{P}_{0|0}+\mathcal{P}_{0|1} + \mathcal{P}_{1|0}+\mathcal{P}_{1|1} \\
    +& \mathcal{P}_{00|00} + \mathcal{P}_{00|11} + \mathcal{P}_{11|00} + \mathcal{P}_{11|11} \\
    -&2\mathcal{P}_{01|01} - 2\mathcal{P}_{01|10} ,
\end{split}
\end{equation}
which simplifies to the compact formulation
\begin{equation}
B = \tilde{\mathcal{P}}_{0} + \tilde{\mathcal{P}}_{00} - 2\tilde{\mathcal{P}}_{01}\geq 0\ ,
\label{eq:THEinequality}
\end{equation}
following the identification of the single- and two-body terms:
\begin{equation}
\begin{split}
    \tilde{\mathcal{P}}_{0}&:= \mathcal{P}_{0|0}+\mathcal{P}_{0|1} + \mathcal{P}_{1|0}+\mathcal{P}_{1|1} \\
    \tilde{\mathcal{P}}_{00}&:= \mathcal{P}_{00|00} + \mathcal{P}_{00|11} + \mathcal{P}_{11|00} + \mathcal{P}_{11|11} \\
    \tilde{\mathcal{P}}_{01}&:= \mathcal{P}_{01|01} + \mathcal{P}_{01|10}  .
\end{split}
\end{equation}}
Note that outcome $2$ does not appear explicitly, as it is eliminated from the normalization of probabilities. Eq.~\eqref{eq:THEinequality} represents a new family of \textit{permutation-invariant Bell inequalities} (PIBIs). As for any other Bell inequality, violation of a PIBI certifies Bell nonlocality. For details on the derivation of Eq.~\eqref{eq:THEinequality} we refer the reader to the Methods, where we also verify that $B\geq 0$ is a \textit{tight} PIBI for arbitrary number of parties $n$. {In turn, in Supplementary Note 1, we explicitly develop Eq.~\eqref{eq:THEinequality} for $n=3$ parties to settle the structure of the proposed PIBI. } \\

Bell inequalities are probed in quantum systems as Bell correlation witnesses, after specifying the measurement settings to maximize the chances of observing nonlocality. Note that} in the multipartite scenario, the cost to infer the values Eq.~\eqref{eq:pij} increases exponentially with the number of parties $n$. However, under some reasonable assumptions, the coarse-grained statistics presented in Eq.~\eqref{eq:Pcol} can be inferred in a scalable manner, from collective observables alone. Such assumptions include the validity of quantum mechanics, an accurate calibration of the measurements, or trusting that the parties are not communicating through unknown channels. In such a case, since we do not follow the strict prescriptions of a Bell experiment, we can no longer claim the detection of Bell nonlocality. Instead, we use the term \emph{Bell correlations} to denote the violation of witnesses constructed from BIs assuming quantum mechanics, and which can be based on collective observables \cite{SchmiedScience2016}. { Nevertheless, it is important to highlight that these very same Bell correlation witnesses certify Bell nonlocality if the involved statistics are inferred from the individual observables, as in a proper Bell test.} Through these witnesses, Bell correlations have been detected successfully in spinor Bose-Einstein condensates (BECs) experiments \cite{SchmiedScience2016} and in cold atomic ensembles \cite{EngelsenPRL2017}. Bell correlations can be detected also in spin chains \cite{Marcin2020, Marcin2022, Marcin2024a,Marcin2024b} from full-body correlations \cite{Jan2022Bell, Niezgoda2020}. Furthermore, refinement of these witnesses enabled a more robust detection \cite{WagnerPRL2017,GuoPRL23}, and the characterization of such correlations through entanglement depth \cite{AloyPRL2019} or Bell correlation depth \cite{BaccariPRA2019}. In this work, we aim at constructing similar Bell correlation witnesses from three-outcome PIBIs, and use them to certify the local dimension of many-body states.   \\ 

{ 
\subsection*{Many-body Bell dimension witnesses from a single-body bound}}
\label{sec:BellDim}

The central contribution of this work is to demonstrate the use of PIBIs as \textit{dimension witnesses}, which are able to certify the local Hilbert space's dimension of the underlying many-body state. When considering Bell inequalities with two dichotomic measurements per party, Jordan lemma ensures their maximal violation can already be achieved by qubits \cite{Jordan1875}. However, for three (or more) outcomes, such as in the case of Eq.~\eqref{eq:THEinequality}, higher dimensional states may be necessary to reach a certain amount of violation. In such a case, the statistics involved in the BI provide a lower bound for the dimension of the local Hilbert space.   \\

For the remainder of this work, we will assume the data Eq.~\eqref{eq:probs} to have a quantum description. In doing so, we associate a local Hilbert space $\mathcal{H}$ to each party, thus expecting that the observed correlations originate from a global quantum state $\hat{\rho}$ on the composite system $\mathcal{H}^{\otimes n}$ according to Born's rule
\begin{equation}
\begin{split}
   p_i(a|x) &= \mathrm{Tr}[\hat{\pi}_{a|x}(i)\hat{\rho}]:= \langle \hat{\pi}_{a|x}(i) \rangle\\
    p_{ij}(ab|xy)  &= \mathrm{Tr}[\hat{\pi}_{a|x}(i)\hat{\pi}_{b|y}(j)\hat{\rho}]:= \langle \hat{\pi}_{a|x}(i)\hat{\pi}_{b|y}(j) \rangle  .
\end{split}
\end{equation}
Here, $\{\hat{\pi}_{a|x}\}_{a\in [r]}$ are generic positive-operator valued measures (POVMs) and the notation $\hat{\pi}_{a|x}(i)$ indicates the application of operator $\hat{\pi}_{a|x}$ on party $i$ with trivial action in the remaining parties.   \\

 Within the quantum framework, the Bell inequality value $B$ is recovered as expectation value of the Bell operator $\hat{\mathcal{B}}$, $B = \langle \hat{\mathcal{B}} \rangle$. Due to PI symmetry of the Bell inequality, the Bell operator is expressed in { terms} of collective operators only, namely operators taking the form $\hat{O} = \sum_{i\in [n]}\hat{o}(i)$, where $\hat{o}(i)$ acts non-trivially only on party $i\in [n]$. Such operators are constructed from the individual POVM elements as 
\begin{equation}
   \left\{ \hat{\Pi}_{a|x}= \sum_{i\in [n]} \hat{\pi}_{a|x}(i) \right\}.
\end{equation}
The Bell operator corresponding to PIBI Eq.~\eqref{eq:THEinequality} is then written as   
\begin{equation}
    \hat{\mathcal{B}} = \underbrace{(\hat{\Pi}_{0|0}-\hat{\Pi}_{1|1})^2 +(\hat{\Pi}_{0|1}-\hat{\Pi}_{1|0})^2}_{:=\hat{\mathcal{B}}_2} + \underbrace{\sum_{i\in [n]} \hat{\beta}(i)}_{\hat{\mathcal{B}}_1} ,
    \label{eq:BellOp}
\end{equation}
where we denote by $\hat{\mathcal{B}}_2$ the quadratic term, which includes two-body all-to-all correlations, and by $\hat{\mathcal{B}}_1$ the remaining collective term. Each addend in $\hat{\mathcal{B}}_1$ is of the form 
\begin{equation}
\begin{split}
    \hat{\beta} &=  \hat{\pi}_{0|0} + \hat{\pi}_{1|1} - (\hat{\pi}_{0|0} - \hat{\pi}_{1|1})^2 \\
 & +\hat{\pi}_{0|1} + \hat{\pi}_{1|0} - (\hat{\pi}_{0|1} - \hat{\pi}_{1|0})^2  .
    \label{eq:Localterm}
\end{split}
\end{equation}
The derivation of Eq.~\eqref{eq:BellOp} can be found in the Methods section. 
Expressing the Bell operator in collective linear, $\hat{\mathcal{B}}_1$, and quadratic, $\hat{\mathcal{B}}_2$, terms, highlights the possibility of using collective measurements to probe PIBI Eq.~\eqref{eq:THEinequality} as a Bell correlation witness. 
As expected, Eq.~\eqref{eq:BellOp} may be inferred from the local measurements $\{\hat{\pi}_{a|x}\}$ in the standard Bell scenario, but also from average of the collective measurements $\hat{\mathcal{B}_1}$ and second moments (i.e. fluctuations) of $(\hat{\Pi}_{0|0}-\hat{\Pi}_{1|1}), (\hat{\Pi}_{0|1}-\hat{\Pi}_{1|0})$ to probe for Bell correlation.      
\\

 In order to derive dimension witnesses, we need to build inequalities fulfilled by any $n$-partite state of local dimension $d = \mathrm{dim}(\mathcal{H})$. To this end, we take the Bell operator for PIBI Eq.~\eqref{eq:THEinequality} and find a lower bound to its expectation value over all single-party POVMs on a $n$-qu\textit{d}it quantum states. Then, if for any unspecified measurement settings on a many-body quantum state we find that such bound is surpassed, one can conclude the corresponding data cannot be explained by $n$ qudits. This implies that at least one of the subsystems needs to have a dimension greater than $d$. Unfortunately, the optimization problem leading to such a bound is outside standard computational capabilities already for few particles $n$. Moreover, convexity is not guaranteed~\cite{Pal2009}. In the following, we propose a relaxation of this problem that is scalable in the number of parties, even more, its complexity does not grow with this parameter.  \\

As mentioned above, the derivation of a dimensionality-constrained bound to a BI is in general extremely demanding in the multipartite setting. However, the structure of the Bell operator Eq.~\eqref{eq:BellOp} enables a scalable relaxation of such a bound. Indeed, we notice that the quadratic part is positive semidefinite (PSD), $\langle \hat{\mathcal{B}}_2\rangle\geq 0$. This observation implies that it is sufficient to bound the linear term $\langle \hat{\mathcal{B}}_1 \rangle$, which leaves us with a single-party problem. Suppose that for a qudit of dimension $d$, the single-party term cannot surpass $\beta_d$, $\langle \hat{\beta} \rangle_d \geq  \beta_d$. Then,
\begin{equation}
   \langle\mathcal{\hat{B}}\rangle\geq n \beta_d \ ,
    \label{eq:DimBell}
\end{equation}
is a proper Bell dimension witness for arbitrary number of subsystems $n$. Regardless of the nature of the measurement settings $\{x\}$ and without relying on trust in their implementation, if the inferred statistics violate Eq.~\eqref{eq:DimBell} one must conclude that qudits of dimension $d$ are insufficient to model the experiment and that at least one subsystem has dimension greater than $d$. 

Next, we propose a generalization of Eq.~\eqref{eq:DimBell} as a number witness. Consider that now we restrict only $n_d\leq n$ parties to have dimension {less or equal than $d$}. Following the same reasoning as in Eq.~\eqref{eq:DimBell}, 
\begin{equation}
\begin{split}
     \langle \hat{\mathcal{B}}\rangle&\geq  n_d\beta_d + (n- n_d)\beta_\infty  \\
     &= n\beta_\infty +  n_d(\beta_d-\beta_\infty):= f(n_d), 
\end{split}
     \label{eq:Ndmax}
\end{equation}
where $\beta_\infty$ is the minimal value of $\langle\hat{\beta} \rangle$ for unbounded local Hilbert space dimension. Note that by construction $\beta_d\geq \beta_\infty$, from which we verify that $f(n_d)$ is an {increasing} function of $n_d$. Indeed, as we restrict the dimension of less parties, larger maximal violation (minimal $ \langle \hat{\mathcal{B}}\rangle$) can be reached. If $\langle \hat{\mathcal{B}}\rangle<f(n_d)$, then the data cannot be explained by restricting $n_d$ parties to have dimension at most $d$. Thus, at least {$n' = n-n_{d}+1  $ }parties must have dimension greater than $d$. The number of certified high dimensional quantum resources from the Bell expectation value is      
\begin{equation}
    n' = \left\lceil\frac{\langle\hat{\mathcal{B}}\rangle-n\beta_d}{\beta_{\infty}-\beta_{d}}\right\rceil
\label{eq:n_d}
\end{equation}
If $\langle \hat{\mathcal{B}}\rangle = n\beta_\infty$ is attainable (i.e. $n\beta$ saturates the Tsirelson bound of the PIBI), then all parties are required to have a dimension greater than $d$.        \\

The optimization of the local term $\langle \hat{\beta} \rangle$ is much simpler than bounding the full Bell operator $\langle \hat{\mathcal{B}} \rangle$. In the next subsection, we elaborate on how to determine the bound $\beta_d$. In particular, we formulate PIBI Eq.~\eqref{eq:THEinequality} as a multi-qutrit witness, i.e. a witness that if violated signals that at least one subsystem is not a qubit. Moreover, beyond Eq.~\eqref{eq:BellOp}, our relaxation can be applied to any PI Bell operator with non-negative collective quadratic term. For instance, when it is decomposed as a sum of squares of collective operators, as in Eq.~\eqref{eq:BellOp}. Many classes of useful PIBIs follow this form \cite{GuillemPRXQuantum}. \\

As detailed in the Methods section, for qubits we obtain a bound $\beta = -1/4$. Thus, we conclude that for any data originating from an $n$-qubit system, the Bell inequality shifted by the qubit bound [cf. Eq.~\eqref{eq:DimBell})]
\begin{equation}
B= \tilde{\mathcal{P}}_{0} + \tilde{\mathcal{P}}_{00} - 2\tilde{\mathcal{P}}_{01}\geq -n/4,
\label{eq:dimwit2}
\end{equation}
is satisfied. Therefore, if such an inequality is violated, one can certify that higher dimensional states (e.g. qutrits) are necessary to describe the inferred correlations. In Methods, we also analyze the maximal violating state for a fixed measurement settings, which resembles a Gaussian superposition of qutrit Dicke states. \\

{In the next section, we further illustrate the usefulness of PIBI Eq.~\eqref{eq:THEinequality} and dimension witness Eq.~\eqref{eq:dimwit2} to detect and characterize nonlocal correlations and multidimensional quantum resources in physically relevant many-body states. To do so, we fix the state and optimize the measurements over certain families to maximize the detection of Bell correlations.} \\

\subsection*{Applications}
\label{sec:applications}

Let us consider a system composed by a number $n$ of three-level parties, that is, a collection of $n$ qutrits. As it turns out, many properties of qubit systems (i.e. spins) are lost in higher dimensions. Here we mention two of them. First, qubit transformations can be understood as rotations, since the SU(2) Lie group is the double cover of the special orthogonal group SO(3). Second, in SU(2) there is only one invariant function of its generators, i.e. one Casimir element; namely the total spin or angular momentum. On the other hand, SU(3) has two Casimir operators, which has consequences for quantum chaos \cite{GnutzmannJPhysA1999} and spin squeezing \cite{YukawaPRA13}. For these (and other) reasons, three-level systems are fundamentally different and much richer than their two-level counterparts.  \\

 The fundamental representation of $\mathfrak{su}(3)$, describing three-level local degrees of freedom $\{\ket{0},\ket{1},\ket{2}\}$, is given by the eight Gell Mann matrices $\{ \hat{\lambda}_a\}$ (see Supplementary Note 5). 
In particular, $\{\hat{\lambda}_0, \hat{\lambda}_1, \hat{\lambda}_2 \}$ constitutes a canonical representative of a $\mathfrak{su}(2)$ subalgebra, as this triad satisfies the spin commutation relations $[\hat{\lambda}_a, \hat{\lambda}_b] = 2i\epsilon_{abc} \hat{\lambda}_a$, where $\epsilon_{abc}$ is the Levi-Civita symbol. 
Following Ref.~\cite{YukawaPRA13}, we denote this subspace as \textit{type-2}. A second spin subalgebra, the so-called \textit{type-1}, is generated by the spin-1 (or dipole) matrices $\{\hat{s}_x, \hat{s}_y, \hat{s}_z \}$. 
This triad can be completed to a basis of $\mathfrak{su}(3)$ by taking certain components of the nematic (or quadrupole) tensor  $\hat{q}_{ab} = \hat{s}_a\hat{s}_b/2 + \mathrm{h.c.}$ (see Supplementary Note 5 for an explicit representation). 
In this context, type-$2$ subspaces can also be written using spin and nematic components, for this reason, they are often called spin-nematic. {In Table~\ref{tab:su3}, we summarize the different subalgebras explored in this work and the basis of operators that we use to express them:

\begin{table}[h!]
    \centering
    \begin{tabular}{|c|c|c|}
    \hline
        $\mathrm{su}(3)$ bases & $\mathfrak{su}(2)$ subalgebra & Type\\
    \hline
  $\{ \hat{\lambda}_a\}_{a\in [8]}$&$\{\hat{\lambda}_0, \hat{\lambda}_1, \hat{\lambda}_2 \}$ & \multirow{3}{*}{2} \\
  \cline{1-2}
\multirow{3}{*}{$\{\hat{s}_a, \hat{q}_{ab} \}_{a,b = \{x,y,z\}}$}  & $\{\hat{s}_x, 2\hat{q}_{yz}, \hat{q}_{zz}-\hat{q}_{yy}\}$&  \\
 & $\{\hat{s}_y, 2\hat{q}_{xz}, \hat{q}_{zz}-q_{yy}  \} $&  \\
  \cline{2-3}
 & $\{\hat{s}_x, \hat{s}_y, \hat{s}_z \}$ & 1 \\
\hline

    \end{tabular}
    \caption{{Examples of $\mathfrak{su}(2)$ subalgebras of $\mathfrak{su}(3)$ and their respective type. }}
    \label{tab:su3}
\end{table}}

As a matter of fact, type-1 and type-2 are the only unitarily inequivalent $\mathfrak{su}(2)$ subalgebras of $\mathfrak{su}(3)$ \cite{YukawaPRA13}. 
To any of these observables, we associate the corresponding collective operators, e.g. $\hat{S}_{x} = \sum_{i\in [n]}\hat{s}_{x}^{(i)}$, $\hat{Q}_{zz} = \sum_{i\in [n]}\hat{q}_{zz}^{(i)}$ etc., which we denote by upper-case letters. Note that these satisfy the same algebra as their single-party analogues.   \\

Multilevel collective measurements are routinely implemented in atomic ensembles, such as in spinor (i.e. multicomponent) Bose Einstein condensates (BECs) \cite{HoPRL98,LawPRL98}. There, the population in each atomic level is usually resolved by splitting the BEC components through an external magnetic field, to then count the number of particles by absorption imaging \cite{chang2004observation}. This effectively corresponds to a Stern-Gerlach-type experiment, which realizes a projective measurement in the $\hat{S}_{z}$ basis. To measure along any other basis, it is possible to implement unitary transformation that bring the desired basis onto $\hat{S}_{z}$ \cite{Giorda_PRA03}. In the concrete, this is implemented experimentally in e.g. alkali-atom hyperfine spin states by rf-pulses and quadratic Zeeman energy shifts \cite{Sau_NJP2010,HamleyNat12}.
We refer the reader to the Supplementary Note 5 for further details.  

{ These peculiarities of SU3 many-body phenomena arise naturally in spinor (i.e. multicomponent) BECs~\cite{HoPRL98,LawPRL98}. Here, we tailor our PIBI to such experiments, although the resulting witnesses may be applied to other collective models as well. The detection of quantum correlations in these systems is relevant to certify the preparation of many-body entangled states with entanglement depth of many thousands of atoms, from the minimal assumptions~\cite{EngelsenPRL2017}. Furthermore, our extension to dimension witnessing enables tests of stronger notions of nonclassicality, specifically allowing us to refute the possibility that the observed correlations originate from an ensemble of qubits. In addition to their fundamental significance, the proposed witness is also of practical interest. As we reveal at the end of this section, metrologically resourceful states maximally violate the witness. Before delving into the three-level case, we highlight how the PIBI Eq.~\eqref{eq:THEinequality} captures physically relevant effective qubit correlations in the form of pseudospin (or spin-nematic) squeezing. To do so, we consider families of measurement settings, and optimize them to detect Bell correlations in spin-nematic squeezed states.


In this first illustration we consider effective two-level projective measurements to test our BIs. We assume all parties having access to the same two measurement settings, both belonging to the type-2 subalgebra spanned by $\{\hat{\lambda}_0, \hat{\lambda}_1, \hat{\lambda}_2 \}$. Then, without loss of generality, we can consider both measurements to lie on $\{\hat{\lambda}_0, \hat{\lambda}_1 \}$ plane with an angle $2\theta$ between each other, namely  
\begin{equation}
\begin{split}
     \label{eq:settings2}
    \hat{m}_0 &= \cos(\theta)\hat{\lambda}_0 +  \sin(\theta)\hat{\lambda}_1  \\
    \hat{m}_1 &= \cos(\theta)\hat{\lambda}_0 -  \sin(\theta)\hat{\lambda}_1   \;.
\end{split}
\end{equation}
The corresponding projectors are $\{\hat{\pi}_{+} = (1+\hat{m})/2,   \hat{\pi}_{-} = (1-\hat{m})/2, \hat{\pi}_{\mathfrak{0}} = \ketbra{2}\}$, and we associate them to the three outputs $\{0,1,2\}$ of the PIBI as $\{+,\mathfrak{0},-\}$ for measurement $0$ and $\{ \mathfrak{0}, -, +\}$ for measurement $1$. 
In this scenario, the expectation value of the Bell operator Eq.~\eqref{eq:BellOp} can be written as a function of $\theta$ and of the data
\begin{equation}
\begin{split}
\label{eq:xyz}
x &:=  \langle \hat{\Lambda}_{0}^2 \rangle /(n-\langle \hat{n}_{2} \rangle) \  \\
y &:= \langle \hat{\Lambda}_1 \rangle/(n-\langle \hat{n}_{2}\rangle) \\
z &:= (n-\langle \hat{n}_{2}\rangle)/n  \;,
\end{split}
\end{equation}
where $\hat{\Lambda}_{a} = \sum_{i\in [n]}\hat{\lambda}_a (i)$ are the collective Gell-Mann matrices and $\hat{n}_2 =\sum_{i\in [n]}\ketbra{2} (i)$ represents the occupation number of level $2$. Minimization of $\langle \hat{\mathcal{B}}\rangle$ over $\theta$ allows us to write $\langle \hat{\mathcal{B}} \rangle\geq 0$ as the Bell correlation witness     
\begin{equation}
    2x + \sqrt{\frac{(z-2)^2}{z^2}-y^2} + \frac{2}{z}-3\geq 0 \;.
\label{eq:spin_nematic_witness}
\end{equation}
Details on the derivation of this result can be found in the Methods section.
In turn, violation of Eq.~\eqref{eq:spin_nematic_witness} reveals Bell nonlocality if the statistics $\{x,y,z\}$ are inferred from a proper Bell scenario, or Bell correlation otherwise.

Let us analyze now the features of the states violating such an inequality. For $z$ constant, the LHS of { Eq.~\eqref{eq:spin_nematic_witness}} is minimized if $x$ (i.e. $\langle\hat{\Lambda}_0^2 \rangle$) is minimal while $y$ (i.e. $|\langle \hat{\Lambda}_1 \rangle|$) is maximal. States with large mean value of $\hat{\Lambda}_1$ and reduced fluctuations of $\hat{\Lambda}_0$ are pseudospin squeezed states. Such class of many-body states display high sensitivity for rotations generated by $\hat{\Lambda}_2$ and thus find application in quantum metrology \cite{PezzeRMP2018}. Their entanglement content has also been characterized extensively \cite{MullerRigat2022, VitaglianoPRL2011, MullerRigat2023,ApellanizNJP2015}. One of the most celebrated entanglement witness for spin squeezing is the Wineland criterion \cite{WinelandPRA1994, SorensenNature2001}, which can be rewritten following the notation introduced in Eq.~\eqref{eq:xyz} as 
\begin{equation}
    \frac{4}{z}-\frac{y^2}{x}\geq 0 \;. 
    \label{eq:Wineland_2}
\end{equation}

Let us now focus on the population measurement $z$. The optimal case (to achieve maximal violation) is when the whole system occupy the two pseudospin levels, so that $\langle \hat{n}_2\rangle = 0$ (or $z = 1$). Then, Eq.~\eqref{eq:spin_nematic_witness} becomes the Bell correlation witness for a 2-outcome PIBI of Ref.~\cite{SchmiedScience2016}. Thus, the new witness we propose can be seen as a generalization to three levels of a known Bell correlation witness. We envision this to be relevant for open systems or imperfect detection, as the third outcome could represent e.g. particle losses or inconclusive measurements.  Along this direction, note that if the number of particles fluctuates from one experimental round to the next, it is possible to introduce the total number operator $\hat{n}$ and replace e.g. $x$ by $\langle \hat{\Lambda}_{0}^2/(\hat{n}-\hat{n}_{2})\rangle$. This is due to the fact that number and pseudospin operators commute, see Ref.~\cite{SchmiedScience2016}. 

Even if the witness we have proposed is tailored to three-level systems, the use of effective two-level measurements (type-$2$) makes it ineffective for dimensionality certification. Indeed, the qubit bound Eq.~\eqref{eq:probquditbound} can never be violated for this choice of measurements. In fact, computing the maximal Bell violation for the proposed measurement settings will never violate $\langle\hat{\mathcal{B}} \rangle\geq -n/4$ and in the thermodynamic limit $n\rightarrow \infty$, the ideal pseudospin-squeezed state saturates such bound. Nonetheless, the witness Eq.~\eqref{eq:spin_nematic_witness} is applicable to detect Bell correlation. We illustrate it below for squeezed states prepared in a spin-1 BECs.  \\

We consider a BEC of $n$ spin-1 atoms with contact interactions subject to an external magnetic field of intensity $B$. In the so-called single mode approximation, the Hamiltonian of the system reads \cite{LawPRL98, Kawaguchi2012}
\begin{equation}
    \label{eq:spin1ham}
    \hat{H} = \frac{c}{2n}\hat{\mathbf{S}}^2 + g\hat{Q}_{zz} ,
\end{equation}
where $\hat{\mathbf{S}} = \sum_{i\in [n]}\hat{\mathbf{s}}{(i)}$ is the collective spin-1 operator, $\hat{Q}_{zz}$ is the collective quadrupole operator, $c$ is the interaction strength and $g\sim B^2$ represents the quadratic Zeemann shift. 

As demonstrated in Ref.~\cite{HamleyNat12}, the spin-mixing dynamics governed by Hamiltonian Eq.~\eqref{eq:spin1ham} induces pseudospin (or spin-nematic) squeezing. In particular, starting from the polar state $\ket{0}^{\otimes n}$, pseudospin squeezing will develop in the type-2 subspace generated by $\{\hat{S}_x, 2\hat{Q}_{yz}, \hat{Q}_{zz} - \hat{Q}_{yy} \}$. To illustrate the detection of Bell correlation in spin-nematic squeezed BEC states, we consider $n = 30$ atoms, $c = -1< 0$ (simulating $^{87}$Rb) and $g = 0.2$. We evolve the polar state $\ket{0}^{\otimes n}$ according to Hamiltonian Eq.~\eqref{eq:spin1ham} for time $t$, which results in $\ket{\Psi(t)} = e^{-it\hat{H}}\ket{0}^{\otimes n}$. For each preparation time $t$, we compute the mean values $\langle \hat{Q}_{zz} + \hat{Q}_{yy} \rangle$, $\langle \hat{Q}_{zz} -\hat{Q}_{xx} +\hat{Q}_{yy}\rangle$ and the fluctuations (second moments) $\{\langle \hat{S}_x^2 \rangle, \langle \hat{Q}_{yz}^2 \rangle$, $\langle \hat{S}_x\hat{Q}_{yz} + \mathrm{h.c.} \rangle \} \}$. Such statistics are sufficient to infer the optimal data corresponding to the squeezed fluctuations $x$, mean (pseudo)spin $y$ and population $z$ entering the witness  Eq.~\eqref{eq:spin_nematic_witness}. See Supplementary Note 6 for details. 

The results of our numerical simulation are shown in Figure~\ref{fig:SpinNematic3D}. The time-evolution dynamics is plotted as a blue line in the $xyz$-space, together with the surfaces representing the Bell correlation witness Eq.~\eqref{eq:spin_nematic_witness} (orange) and the Wineland entanglement criterion Eq.~\eqref{eq:Wineland_2} (green). 
At $t=0$, the polar state $\ket{0}^{\otimes n}$ appears in the top-right corner, with coordinates $(x,y,z)=(1,1,1)$. As time progresses, $x$ is reduced, i.e. spin-nematic squeezing is generated. This results very quickly in the violation of Eq.~\eqref{eq:Wineland_2}, which signals entanglement among the particles in the systems. For longer evolution times the correlations get so strong that Eq.~\eqref{eq:spin_nematic_witness} is also violated, indicating the presence of Bell correlations. Ultimately, even longer evolution times results in over-squeezed states with small polarizations (reduced first moments of the collective spin), for which the considered witnesses are inadequate to detect correlations. { Figure~\ref{fig:SpinNematic3D} also transparently illustrates, as announced earlier, how Bell nonlocality is a much more stringent resource than quantum entanglement, as witnessed by Eq.~\eqref{eq:Wineland_2}. The violation of the Bell inequality (in the form of the witness) by spin-nematic squeezed states represents one of the strongest tests of nonclassicality, beyond quantum entanglement. Moreover, its appeal to quantum-enhanced metrological applications extends the connections between metrological usefulness and Bell nonlocality, which have recently attracted significant interest~\cite{FroewisPRAR2019, Niezgoda2021}. In the following section, we verify how the same experiment violates the Bell dimension witness by fully exploiting all three hyperfine levels.  } \\

\begin{figure}[h]
\centering
\includegraphics[width=0.9\linewidth]{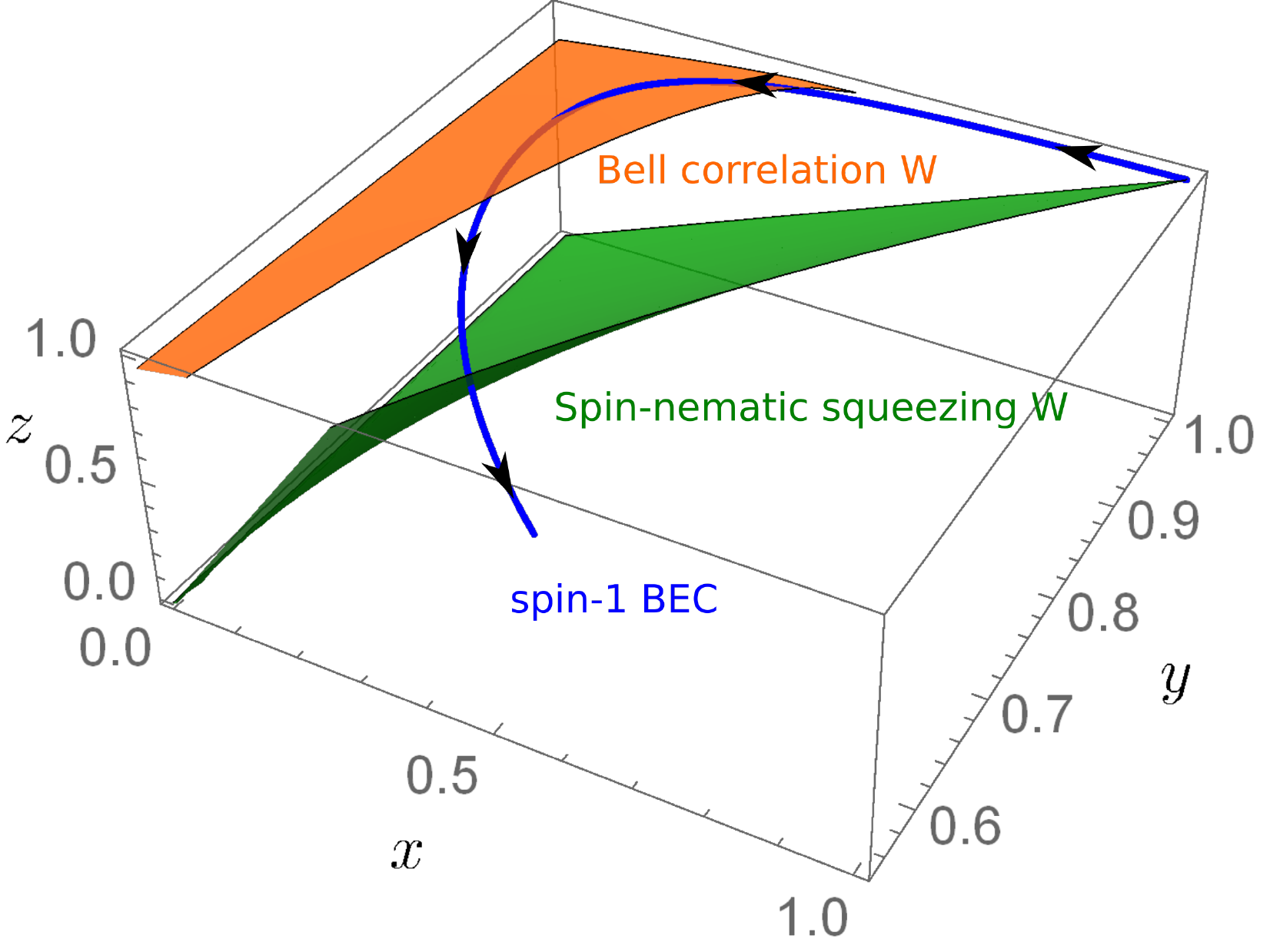}
\caption{Comparison of the spin-nematic witness Eq.~\eqref{eq:Wineland_2} (green surface) and the Bell correlation witness Eq.~\eqref{eq:spin_nematic_witness} (orange surface). Data above such surfaces indicates a violation of the corresponding witness. The solid blue line represents the data obtained from evolving a polar state $\ket{0}^{\otimes n}$ of $n= 30$ spin-1 particles according to Hamiltonian Eq.~\eqref{eq:spin1ham}. }
\label{fig:SpinNematic3D}
\end{figure}


For the next illustration, we will consider instead spin-1 projections to construct the measurement strategy. As before, we assume all parties having access to the same two measurement settings, both belonging to the type-1 subalgebra spanned by $\{\hat{s}_x, \hat{s}_y, \hat{s}_z \}$. As above, without loss of generality, we can consider both measurements to lie on $\{\hat{s}_x, \hat{s}_y \}$ plane with an angle $2\theta$ between each other, namely 
\begin{equation}
\begin{split}
\hat{m}_0 &= \cos(\theta)\hat{s}_x + \sin(\theta)\hat{s}_y \\
\hat{m}_1 &= \cos(\theta)\hat{s}_x - \sin(\theta)\hat{s}_y \;.
\end{split}
\end{equation}
We associate output $0$ with the projector onto spin projection $+1$ and likewise output $1$ for spin $-1$ for both measurements, i.e., $\hat{\pi}_{0|0} = (\hat{m}^2_0 + \hat{m}_0)/2,\hat{\pi}_{0|1} = (\hat{m}^2_1 + \hat{m}_1)/2, \hat{\pi}_{1|0} = (\hat{m}^2_0 - \hat{m}_0)/2, \hat{\pi}_{1|1} = (\hat{m}^2_1 - \hat{m}_1)/2 $. Considering this choice of settings, the Bell operator $\hat{\mathcal{B}}$ is a function of $\theta$ and of the data $(x,y,z)$ that we redefine here as
\begin{equation}
\begin{split}
   x &= \langle \hat{S}_x^2 \rangle/n \\
   y &= \langle 3\hat{Q}_{zz} + \hat{Q}_{xx} -8\hat{Q}_{xy}^2-2n\mathbb{I}\rangle/n\\
   z &= \langle \hat{Q}_{yy}\rangle/n .
\end{split}
\end{equation}

Next, we proceed to minimize the Bell expectation value $\langle \hat{\mathcal{B}}(\theta)\rangle$ over $\theta$, which results now in the witness
\begin{equation}
   2 x- y+3z +2\sqrt{(2z-\beta)(y+z)}\geq 0 \;,
    \label{eq:type1BellCor}
\end{equation}
where $\beta$ is the considered bound, i.e. $\langle \hat{\mathcal{B}} \rangle\geq n\beta$. For $\beta = 0$, Eq.~\eqref{eq:type1BellCor} serves as a Bell correlation witness. On the other hand, for $\beta = -1/4$ it serves as a Bell dimension witness for qutrits. If the latter is violated, it indicates that at least one subsystem has dimension greater than two. Moreover, if it is violated with $\beta = -1/4 +(1-n')/(4n)$ [cf. Eq.~\eqref{eq:Ndmax}], it indicates that $n'$ qutrits are necessary to reproduce the statistics $\{x,y,z\}$. Remarkably, in the limit of a perfectly squeezed state, $\{x= 0, y =2, z = 0\}$, this data can by generated only if all parties $n$ are qutrits. The derivation of Eq.~\eqref{eq:type1BellCor} can be found in the Methods section. Let us examine closer Eq.~\eqref{eq:type1BellCor}. The states which maximally violate the witness display simultaneous squeezing in two non-commuting observables, in such case, $\hat{S}_x$ and $\hat{Q}_{xy}$. The practical witness Eq.~\eqref{eq:type1BellCor}, in turn, serves to optimally certify the Bell correlation and qutrit resource content from the squeezed components of such metrologically useful states. \\ 

We finish the paper by demonstrating experimentally-accessible detection of genuine qutrit correlations in spin-1 BEC. We consider the same scenario as the previous example, namely evolution of the polar state $\ket{0}^{\otimes n}$ under the spin-mixing Hamiltonian Eq.~\eqref{eq:spin1ham}. Here, we will use the fact that such an evolution, do not only generate spin-nematic squeezing in the $\{\hat{S}_x, 2\hat{Q}_{yz}, \hat{Q}_{zz} - \hat{Q}_{yy} \}$ subspace. Indeed, the type-2 subspace generated by $\{\hat{S}_y, 2\hat{Q}_{xz}, \hat{Q}_{xx}-\hat{Q}_{zz}\}$ can be found simultaneously squeezed. Thus, such states can be exploited for multiparameter estimation tasks \cite{Fadel_NJP2023}, as it was recently demonstrated experimentally \cite{cao2023}. { Interestingly, this double phase estimation can be implemented jointly through protocols involving simultaneous readout of non-commuting observables (see also Ref.~\cite{kunkel2019simultaneous}), which, as reported, lead to below shot-noise sensitivity with respect to both collective components.} As already announced, the witness Eq.~\eqref{eq:type1BellCor} is maximally violated for states with reduced fluctuations of $\hat{S}_x$ and $\hat{Q}_{xy}$. We transform locally the state such that the previous data correspond to the two squeezed orientations and optimize the witness accordingly (see Supplementary Note 6). By doing so, we are able to certify not only Bell correlations but quantum resources of high dimensionality signaled by the violation of the Bell dimension witness. The results are presented in Figure~\ref{fig:3Cor3D}.      

\begin{figure}[h]
\centering
\includegraphics[width=0.9\linewidth]{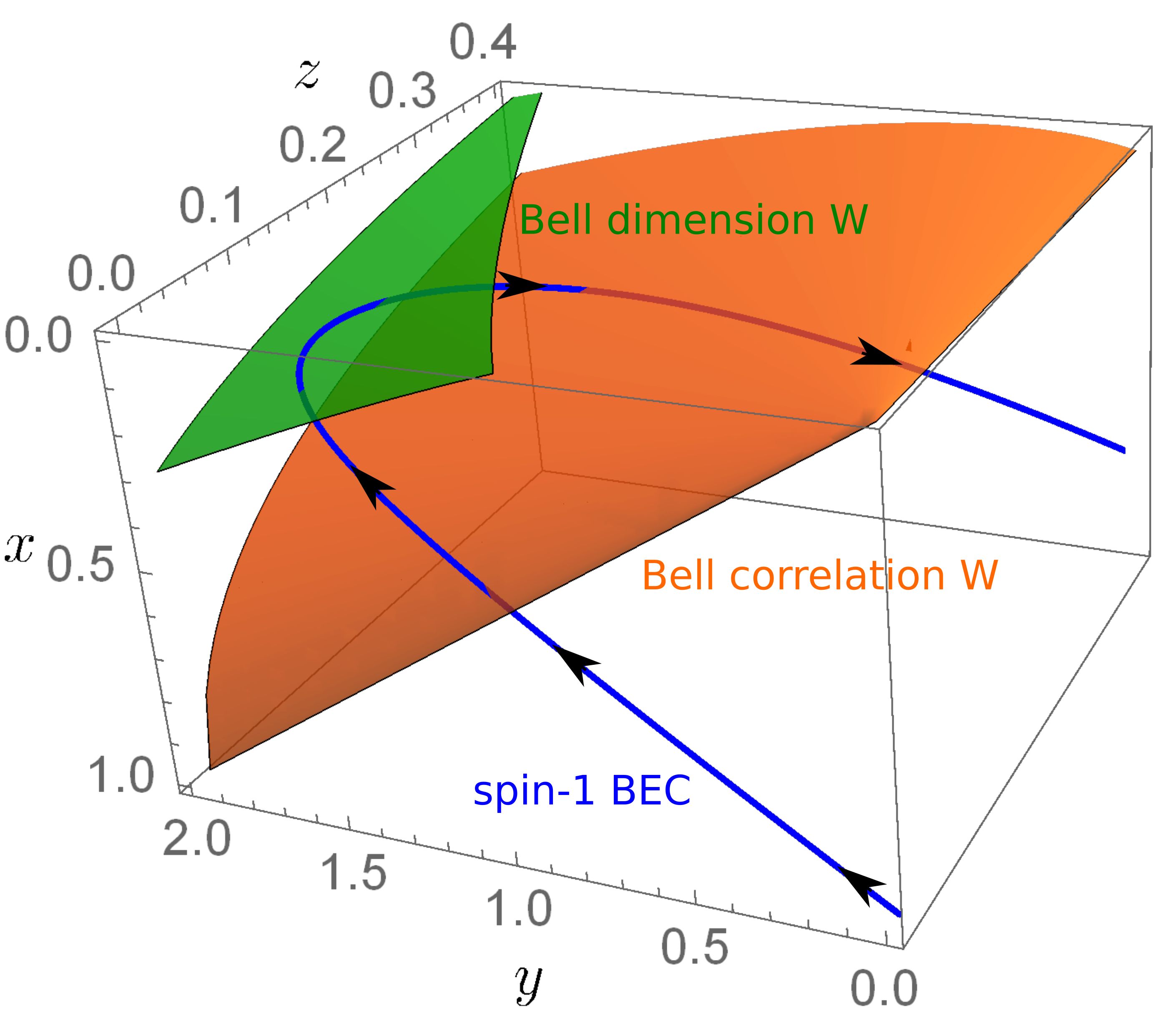}
\caption{Comparison of the dimension witness Eq.~\eqref{eq:type1BellCor} for $\beta = 0$ (orange surface) and $\beta = -1/4$ (green surface). Data above such surfaces indicates a violation of the corresponding witness. In the specific, data above the orange surface signals Bell correlation while data above the green surface reflects expectation values that cannot be reproduced by multi-qubit systems. The solid blue line represents the data obtained from spin-nematic squeezed states as prepared via the spin-1 BEC evolution governed by Hamiltonian Eq.~\eqref{eq:spin1ham} starting from the polar state $\ket{0}^{\otimes n}$ for $n=70$ atoms.}
\label{fig:3Cor3D}
\end{figure}

In Figure~\ref{fig:3Cor3D}, we illustrate as surfaces in the $xyz$-space the witnesses obtained from Eq.~\eqref{eq:type1BellCor} for $\beta=0$ (orange) and $\beta=-1/4$ (green). The blue line corresponds to the data derived from the spin-1 BEC for $n=70$ atoms. The initial polar state $\ket{0}^{\otimes n}$ displays statistics $(1,0,0)$. Time evolution generates simultaneous squeezing in two pseudospin subspaces, which eventually violates the SU3 Bell correlation witness Eq.~\eqref{eq:type1BellCor}, thus signalling Bell nonlocality. Next, sufficient squeezing is developed to violate the Bell dimension witness. At such time interval, we are able to certify high dimensional states as qubits are insufficient to describe the corresponding statistics. Finally, we recover the expected behaviour at longer times, when the state becomes non-Gaussian and second-order moments fail to capture these quantum effects.  \\

\section*{DISCUSSION}
\label{sec:conclusions}

In this work, we have introduced a scalable permutationally invariant Bell inequality based on one- and two-body correlation functions of three-outcome measurements. Such a Bell inequality is suited to probe quantum correlations in three-level (or spin-1) system of arbitrary number of particles, since it can be converted into a Bell correlation witness that depends on experimentally accessible collective observables. {Note that these experiments require trusted measurement assumptions and do not constitute a loophole-free Bell test.} 

From a modification of this witness, we propose a method to derive Bell dimension witnesses, namely criteria that signal the impossibility of a given statistics to originate from measurements on an $n$-body state of some single-particle Hilbert space dimension $d$. The method we propose is based on computing scalable bounds to the maximal violation achievable with dimension-constraint single-particle Hilbert spaces. In particular, we focus on the qubit case, $d=2$, and as a result, we establish a Bell dimension witness tailored to systems of arbitrary size. Then, we moved to discuss the relevance of the witness by characterizing the quantum states and measurements that violate it the most. These include Gaussian superpositions of qutrit Dicke states that can appear as ground states of SU(3) Hamiltonians~\cite{LipkinNucPhys1965,GnutzmannJPhysA1999} and spin-nematic squeezed states, as the one demonstrated in spin-1 Bose Einstein condensates~\cite{HamleyNat12}. However, despite the fact that these correlations originate in spin-1 systems, they can effectively be described by qubits \cite{KitzingerPRA21}. We show that states that do violate the dimension witness, and thus are certified to be genuine three-level nonlocal, correspond to SU(3) states where two effective qubit subspaces appear simultaneously squeezed. These states are also prepared in spin-1 BECs and can be exploited for multiparameter estimation protocols, as shown very recently~\cite{cao2023,Fadel_NJP2023}. It is also worth noting how beyond the specific inequality addressed in this work, our method can be applied to other multilevel multipartite Bell inequalities that have been proposed in the past \cite{GuillemPRXQuantum}.

Notably, our approach can be extended to certify the number of high-dimensional parties in a many-body system. In this regard, we demonstrate that, in the thermodynamic limit, the squeezed correlations described above that maximally violate the PIBI require the system to be composed entirely of qutrits. These results shed light on the fundamental role of high-dimensional quantum resources in the macroscopic scale and their application for multiparameter estimation purposes.

In summary, our work paves the way to test Bell correlations in multipartite spin-1 systems and to use many-outcome Bell inequalities as dimension witnesses. 
These can be adopted experimentally to lower-bound the number of high-dimensional particles in a many-body system from the presence of nonlocal correlations. Specifically, certain paradigmatic models for condensed-matter physics, such as the bilinear-biquadratic spin-1 chain, display spin-nematic correlations as signatures of topological phases~\cite{Lauchli2006}. It would be highly beneficial to study these quantum phases through the lens of high-dimensional quantum resources with the certification tools introduced in this work. \\

\section*{METHODS}
\subsection*{Three outcome permutation invariant Bell inequality}

We provide in this section details on the derivation of permutationally invariant Bell inequalities (PIBIs), as well as of their classical bound.  \\

The polytope of classical correlations may be constructed from its extremal points, i.e. from local deterministic strategies (LDS)~\cite{Pitowski1989}.  LDS are rules which assign to each measurement setting a deterministic outcome. In the generic scenario, the number of LDS grows exponentially with $n$. Conversely, if PI is exploited, these can be characterized only by a set of non-negative integers $\{c_\mathbf{\mathbf{a}}\}$, each indicating the number of parties which exhibit outcomes $\mathbf{a} = a_0a_1...a_{k-1}$ when measuring setting $\{0,1,..,k-1\}=[k]$. For instance, in the Bell scenario that we primarily address, $k=2,r = 3$, one has $\{c_{00}, c_{01},c_{02}, c_{10}, c_{11}, c_{12}, c_{20}, c_{21}, c_{22} \}$. Note that $c_\mathbf{a}\geq 0$ and $\sum_{\mathbf{a}}c_\mathbf{a} = n$. It is important to acknowledge how permutation invariance reduces the number of LDS from $r^{kn}$ to ${n+r-1 \choose k-1}\sim \order{(n + r -1)^{(k-1)}/(k-1)!}$.

The collective probabilities defined in the main text decompose under LDS as 
\begin{equation}
   \label{eq:LDS}
\begin{split}
    \mathcal{P}_{a|x} &= \sum_{\mathbf{r}\cdot\mathbf{x} = a}c_{\mathbf{r}} \\
\mathcal{P}_{ab|xy} &=\sum_{\mathbf{r}\cdot\mathbf{x} = a}c_{\mathbf{r}}\sum_{{ \mathbf{r}'\cdot\mathbf{y} = b}}c_{\mathbf{r}'} - \sum_{\substack{\mathbf{r}\cdot\mathbf{x} = a \\ \mathbf{r}\cdot\mathbf{y} = b}}c_{\mathbf{r}} \;,
\end{split}
\end{equation}
where the notation $\mathbf{r}\cdot\mathbf{x} = a$ indicates all LDS $\{c_{\mathbf{r}}\}$ with $\mathbf{r}$ displaying $a\in [r]$ at the position $x\in [k]$. For example, in the above Bell scenario, $\mathcal{P}_{0|0} = c_{00} + c_{01} + c_{02}$ amounts to all parties with outcome $a=0$ if $x=0$ is measured. 

For sufficiently small number of parties $n$, one is able to enumerate the vertices of the local polytope in the space of PI correlations $\{\mathcal{P}_{a|x},  \mathcal{P}_{ab|xy}\}$. The convex hull of the vertices defines the polytope. An equivalent description of a geometrical object is through intersection of half-planes, i.e. Bell inequalities. Switching between these two representations is a linear program, which is routinely solved via well-established algorithms \cite{fukuda1997cdd}.   \\

In order to improve further the scalability of the vertex enumeration problem, we introduce an additional symmetrization over the inputs and outputs labels. In the $(n,2,3)$ scenario, such simplification reduces the dimension of the correlation space to the five quantities
\begin{equation}
\begin{split}
\tilde{\mathcal{P}}_{0} &= \mathcal{P}_{0|0}+P_{0|1} + P_{1|0}+P_{1|1} \\
\tilde{\mathcal{P}}_{00}  &= \mathcal{P}_{00|00} + \mathcal{P}_{00|11} + \mathcal{P}_{11|00} + \mathcal{P}_{11|11}  \\
\tilde{\mathcal{P}}_{10}  &=  \mathcal{P}_{00|01} + \mathcal{P}_{11|01} \\
\tilde{\mathcal{P}}_{11}  &=  \mathcal{P}_{01|00} + \mathcal{P}_{01|11} \\
\tilde{\mathcal{P}}_{01} &=  \mathcal{P}_{01|01} + \mathcal{P}_{01|10}
\end{split}.
\label{eq:tilde_vars}
\end{equation}
Note that we disregarded outcome $2$ without loss of generality by virtue of the non-signaling principle and normalization of the marginals \cite{UsInPrep_theta}. The symmetry considered here is motivated by the fact that input and outcome labeling is a free choice, and thus arbitrary.
{ In a nutshell, the symmetrization assigns the same coefficients of the BI ``$\alpha$'' to correlations that are related by permutations of inputs and outputs. Let us first consider the terms of the form $\mathcal{P}_{aa|xx}$. Permutation of outputs identifies $\mathcal{P}_{00|00}, \mathcal{P}_{11|00}$ and  $\mathcal{P}_{00|11}, \mathcal{P}_{11|11}$. In turn, permutation of inputs identifies e.g.  $\mathcal{P}_{00|00}, \mathcal{P}_{00|11}$, which prompt us to group the four correlators under the same symbol $\tilde{\mathcal{P}}_{00}$ The case $\mathcal{P}_{ab|xy}$ $a\neq b, x\neq y$  (term $  \tilde{\mathcal{P}}_{01}$) is more intricate. By permutation invariance between any pair of parties, we observe that $\mathcal{P}_{ab|xy}$ and $\mathcal{P}_{ba|yx}$ coincide. And by symmetrization of inputs and outputs we group them with  $\mathcal{P}_{ab|yx}, \mathcal{P}_{ba|xy}$. For further details, we encourage the interested reader to consult Ref.~\cite{Aloy_Lopez2023}.    }

We proceed to fully characterize PIBIs in this reduced space, where they can be parametrize by the coefficients $\vec{\alpha} = (\alpha_0, \alpha_1, \alpha_2, \alpha_3, \alpha_4)$ and the classical bound $\beta_c$ as    
\begin{equation}\label{eq:suppPIBI5}
    B =  \alpha_0 \tilde{\mathcal{P}}_{0}  + \alpha_1 \tilde{\mathcal{P}}_{00} + \alpha_2 \tilde{\mathcal{P}}_{10} +  \alpha_3 \tilde{\mathcal{P}}_{11} + \alpha_4 \tilde{\mathcal{P}}_{01} - \beta_c\geq 0 \;.
\end{equation}
Such coefficients are in general dependent on the number of particles $n$. Then, we select those PIBIs with constant coefficients, which suggests scalable inequalities that could be valid for any $n$. Among the inequalities found by this approach, we identified 
\begin{equation}
B = \tilde{\mathcal{P}}_{0} + \tilde{\mathcal{P}}_{00} - 2\tilde{\mathcal{P}}_{01}\geq 0 \;, 
\label{eq:THEinequalityAPP}
\end{equation} 
coming from $\vec{\alpha} = (1,1,0,0,-2) $ and $\beta_c = 0$, as presented in the main text. We point the reader to Ref.~\cite{UsInPrep_theta} for a pedagogic introduction to the characterization of such projected polytopes and for more PIBIs of the form Eq.~\eqref{eq:suppPIBI5}. 

Remarkably, as we are going to prove in the following, Eq.~\eqref{eq:THEinequalityAPP} is a facet of the projected polytope generated by Eq.~\eqref{eq:tilde_vars} for any $n$ divisible by $12$.  \\

 To prove that the classical bound is tight, it is sufficient to verify that it can be saturated by LDS. These are characterized by the number of parties which exhibit outcomes $(a,b)$ while measuring $(0,1)$, namely by $\{c_{ab}\geq 0\}$ with $\sum_{a,b\in [3]}c_{ab} = n$. If we define $\mathbf{c} = (c_{00}, c_{11},c_{02}, c_{20}, c_{12}, c_{21}) $ then, after subtracting $ 2(c_{10} + c_{01}) \geq 0$, we have
\begin{equation}
    B_{\mathrm{LDS}} \geq  \mathbf{c}^T
    \underbrace{
    \begin{pmatrix}
    2 & -2 & 1 & 1& -1 & -1 \\
    -2 & 2& -1 & -1 & 1 & 1 \\
    1 & -1 & 1 & 0 & 0 & -1 \\
    1 & -1 & 0 & 1 & -1 & 0 \\
    -1 & 1 & 0 & -1 & 1 & 0 \\
    -1 & 1 & -1 & 0 & 0 & 1
    \end{pmatrix}}_{:= \mathcal{I}_{\rm LDS}} 
    \mathbf{c} \;.
    \label{eq:PSD_classical}
\end{equation}
\vspace{5cm}
It is easy to verify that $\mathcal{I}_{\rm LDS}$ is positive-semidefinite (PSD), i.e. $\mathcal{I}_{\rm LDS}\succeq 0$, therefore $B_{\mathrm{LDS}}\geq 0$. In particular, $c_{22}=n$ gives $B_{\mathrm{LDS}} =0$, which confirms that the classical bound is tight, i.e. it cannot be improved for any number of parties $n$. An alternative proof for the classical bound can be found in Ref.~\cite{UsInPrep}.

To show under which conditions the inequality is a facet of the projected polytope in the $5$-dimensional space generated by Eq.~\eqref{eq:tilde_vars}, we have to exhibit 5 linearly independent points $\vec{v} = (\tilde{P}_0, \tilde{P}_{00}, \tilde{P}_{10},\tilde{P}_{11}, \tilde{P}_{01})$ evaluated by a LDS $\mathbf{c}, c_{01}, c_{10}, c_{22}$ such that the inequality is saturated, $B = 0$. If this is the case, we verify that the inequality describes a facet of the polytope as a supporting hyperplane of maximal dimension. The LDS strategy $c_{22} = n$ already provides the first point $\vec{v} = (0,0,0,0,0)$ and tells us that the hyperplane contains the origin of coordinates. The remaining 4 points are found by examining the kernel of the matrix $\mathcal{I}_{\rm LDS}$  as follows. The kernel of $\mathcal{I}_{\rm LDS}$ is generated by the vectors $\mathbf{c}_1 = (0, 0, 1, 0, 0, 1) , \mathbf{c}_2 =(1, 0, -1, 0, 1, 0), \mathbf{c}_3 = (-1, 0, 1, 1, 0, 0), \mathbf{c}_4 = (1, 1, 0, 0, 0, 0)$. Thus, a general LDS giving $B=0$ can be parametrized as follows $c_{01} = c_{10} = c_{22}=0, \mathbf{c}(x,y,z) = n(x\mathbf{c}_1 + y\mathbf{c}_2 + z\mathbf{c}_3 + (1 - 2 x - y - z)\mathbf{c}_4/2)$. Note that by construction $\mathrm{sum}(\mathbf{c}) = n$ as it should. The subset of $(x,y,z)$ such that $\mathbf{c}(x,y,z)\geq 0$ component-wise forms a polytope in $\mathbb{R}^3$. In Table~\ref{tab:LDS} we enumerate their vertices, the LDS each represents, and the corresponding supporting point $\vec{v}$.  
\begin{widetext}
\begin{center}
\begin{table}
\begin{tabular}{ |c|c|c| } 
 \hline
$(x,y,z)$ & LDS & $\vec{v} = (\tilde{P}_0, \tilde{P}_{00}, \tilde{P}_{10},\tilde{P}_{11}, \tilde{P}_{01})$ \\ 
\hline
$(0,0,0)$ & $c_{00} = c_{11} = n/2$ & $(2 n, -2 n + n^2, -n + n^2/2, n^2/2, n^2/2)$ \\ 
 $(1/2,0,0)$ & $c_{02} = c_{21} = n/2$ & $(n, -n + n^2/2, 0, 0, n^2/4)$ \\ 

  $(0,0,1/3)$ & $c_{11} = c_{02} = c_{20} = n/3$ & $(4 n/3, -4 n/3 + 4 n^2/9, -n/3 + 2 n^2/9, 2 n^2/9, 
 2 n^2/9)$ \\

 $(1/3,1/3,0) $ & $c_{00} = c_{12} = c_{21} = n/3$  &  `` \hspace{3cm} "\\
  $(0,1/4,1/4)$  & $c_{00} = c_{11} = c_{20} = c_{12} = n/4$  &  $(3 n/2, -3 n/2 + 5 n^2/8, -n/2 + n^2/4, n^2/4, 5 n^2/16)$\\
 \hline
\end{tabular}
\caption{Local deterministic strategies.}
\label{tab:LDS}
\end{table}
\end{center}

\end{widetext}

To finish the proof, we verify that the orthogonal complement of the 4 vectors $\vec{v}$ is indeed $\vec{\alpha} = (1,1,0,0,-2)$. The LDS corresponding to the supporting planes are well defined for $n$ is divisible by $2,3,4$.

\subsection*{Derivation of the Bell operator}

The quantities defined in Eq.~\eqref{eq:Pcol} become in the collective picture
\begin{equation}
\begin{split}
    \mathcal{P}_{a|x} &= \langle \hat{\Pi}_{a|x} \rangle  \\
    \mathcal{P}_{ab|xy} & = \langle \{\hat{\Pi}_{a|x},\hat{\Pi}_{b|y}\}/2  -\hat{\Pi}_{ab|xy}  \rangle \ ,
\end{split}
\end{equation}
where,
\begin{equation}
\begin{split}
    \hat{\Pi}_{a|x} &= \sum_{i\in [n]}\hat{\pi}_{i}(a|x) \\ 
    \hat{\Pi}_{ab|xy} &= \sum_{i\in [n]}\{\hat{\pi}_{i}(a|x),\hat{\pi}_{i}(b|y) \}/2 \ ,
\end{split}
\end{equation}
are collective operators and $\{\cdot,\cdot \}$ is the anticommutator. 
From these, we derive the symmetrized one and two-body terms 
\begin{equation}
\begin{split}
   \tilde{\mathcal{P}}_0 &= \langle \hat{\Pi}_{0|0}+\hat{\Pi}_{0|1}+\hat{\Pi}_{1|0}+\hat{\Pi}_{1|1}\rangle \\
   \tilde{\mathcal{P}}_{00} &= \langle\hat{\Pi}_{0|0}^2 - \hat{\Pi}_{00|00} + \hat{\Pi}_{0|1}^2 - \hat{\Pi}_{00|11}  \\
   &+\hat{\Pi}_{1|0}^2 - \hat{\Pi}_{11|00}+\hat{\Pi}_{1|1}^2 - \hat{\Pi}_{11|11}\rangle \\
   -2\tilde{\mathcal{P}}_{01} &=\langle \hat{\Pi}_{01|01}+\hat{\Pi}_{01|10} - \hat{\Pi}_{0|0}\hat{\Pi}_{1|1} \\
   &-\hat{\Pi}_{1|1}\hat{\Pi}_{0|0} - \hat{\Pi}_{0|1}\hat{\Pi}_{1|0}-\hat{\Pi}_{1|0}\hat{\Pi}_{0|1}\rangle \ .
\end{split}
\end{equation}
Summing all contributions yields $\langle \hat{\mathcal{B}}\rangle$ with $\hat{\mathcal{B}}$ the Bell operator 
\begin{align}
  \hat{\mathcal{B}}=(\hat{\Pi}_{0|0}-\hat{\Pi}_{1|1})^2+(\hat{\Pi}_{0|1}-\hat{\Pi}_{1|0})^2+\hat{\mathcal{B}}_1\ ,
\end{align}
where,
\begin{equation}
\begin{split}
    \hat{\mathcal{B}}_1 &=\hat{\Pi}_{0|0}+\hat{\Pi}_{0|1}+\hat{\Pi}_{1|0}+\hat{\Pi}_{1|1}  \\
    &-\hat{\Pi}_{00|00} - \hat{\Pi}_{00|11} -\hat{\Pi}_{11|00} - \hat{\Pi}_{11|11} \\
    & +2(\hat{\Pi}_{01|01} +\hat{\Pi}_{01|10}) \ .
\end{split}
\end{equation}
If the local measurements are projective, $\hat{\pi}_{a|x}\hat{\pi}_{b|x} = \delta_{ab}\hat{\pi}_{b|x}$, the one-body term further simplifies to
\begin{equation}
    \hat{\mathcal{B}}_1 = 2(\hat{\Pi}_{01|01} + \hat{\Pi}_{01|10}) \ .
\end{equation}

\subsection*{Computation of the dimension-constrained bounds}

The local operator $\hat{\beta}$ Eq.~\eqref{eq:Localterm} can be interpreted as a polynomial $p$ of degree two in the non-commuting variables corresponding to the measurement operators $\hat{\pi} = \{\hat{\pi}_{a|x}\}$, namely $\hat{\beta} = p(\hat{\pi})$. 
The optimization that we have to tackle is a dimension-constrained polynomial optimization problem 
\begin{equation}
    \beta_d := \begin{array}{ccc} \min_{\hat{\mathbf{\pi}},\ket{\psi}\in \mathbb{C}^d}& \bra{\psi} p(\hat{\pi}) \ket{\psi}  \\
    \mbox{s.t.}& \hat{\pi}_{a|x}\succeq 0, \sum_{a}\hat{\pi}_{a|x} = \mathbb{1}
    \end{array} \ .
    \label{eq:probquditbound}
\end{equation}
This class of problems have been studied extensively in the past, also to bound BIs over dimension constraints \cite{navascues2015bounding, navascuesPRA2015}. However, to the best of our knowledge, none of these previous works are applicable to multipartite scenarios because of the exponential scaling of the number of correlators to be considered.
Here, we highlight how solving a single-party version of such problem may be sufficient to derive useful Bell dimension witnesses.

The polynomial $p(\hat{\pi})$ still contains quadratic terms, which makes its optimization nontrivial and prompt us to resort to numerical methods. In the following, we propose two approaches to evaluate bound Eq.~\eqref{eq:probquditbound}, and apply them to PIBI Eq.~\eqref{eq:THEinequality} for $d=2$ (qubits).  \\


The first approach we propose is of variational nature, meaning that we directly express the problem Eq.~\eqref{eq:probquditbound} as the following unconstrained optimization. First, we introduce a parametrization of the local $r$-outcome qudit POVMs  $\{\hat{\pi}_{a|x}(\boldsymbol{\theta})\}$ (see Supplementary Note 2). Then, we numerically minimize $\bra{0}p(\hat{\pi})\ket{0}$ over $\boldsymbol{\theta}$ to obtain as a solution $\tilde{\beta}_d \geq \beta_d$.
For the case of Eq.~\eqref{eq:Localterm}, the optimization gives $\tilde{\beta}_2 = -1/4$ for qubits and $\tilde{\beta}_3 = -1/2$ for qutrits. 

In addition, the algorithm also returns the optimal measurement settings, which appear to be the projectors $\hat{\pi}_{a|x} = \ketbra{a,x}$. 
For $d=2$, after a suitable local unitary transformation, we can bring setting $x=0$ to the computational basis, $\{\hat{\pi}_{0|0} = \ketbra{0},\hat{\pi}_{1|0} = 0,  \hat{\pi}_{2|0} = \ketbra{2}\}$. Then, the optimal measurement $x=1$ is described by $\hat{\pi}_{0|1} = 0$ and  
\begin{align}
\begin{split}
\label{eq:qubitoptproj}
\ket{1,1} &= \frac{1}{2}\ket{0} + \frac{\sqrt{3}}{2}\ket{2} \\
\ket{2,1} &= \frac{\sqrt{3}}{2}\ket{0} - \frac{1}{2}\ket{2} ,
\end{split}
\end{align}
Notice that both measurement settings are effectively two-level, as they act only on $\{\ket{0},\ket{2}\}$. 

For $d=3$, as done above, we can write setting $x=0$ in the computational basis, $\{\hat{\pi}_{0|0} = \ketbra{0},\hat{\pi}_{1|0} = \ketbra{1},  \hat{\pi}_{2|0} = \ketbra{2}{2}\}$, which results in the optimal measurement $x=1$ to be
\begin{equation}
\begin{split}
\label{eq:qutritoptproj}
\ket{0,1} &= \frac{1}{2}\ket{0} + \frac{1}{2}\ket{1} + \frac{1}{\sqrt{2}}\ket{2}  \\
\ket{1,1} &= \frac{1}{2}\ket{0} + \frac{1}{2}\ket{1} - \frac{1}{\sqrt{2}}\ket{2} \\
\ket{2,1} &=     \frac{1}{\sqrt{2}}\ket{0} -   \frac{1}{\sqrt{2}}\ket{1}                    \ .
\end{split} 
\end{equation}
Consider that $\{\ket{0},\ket{1},\ket{2}\}$ have well-defined spin-1 along the $z$ direction, $\hat{s}_z$ (see Supplementary Note 5 for its explicit representation), corresponding to $\{1,-1,0\}$ respectively. Then, the states $\{\ket{0,1},\ket{1,1},\ket{2,1}\}$ are eigenstates of the $\hat{s}_x$ observable with eigenvalues $\{1,-1 ,0\}$, respectively. Thus, one may use single-party spin-1 observables, $\hat{s}_z$ and $\hat{s}_x$, to probe for Bell correlations and for their dimensionality with PIBI Eq.~\eqref{eq:THEinequality}, cf. Applications section.   \\

The problem Eq.~\eqref{eq:probquditbound} addressed with the variational approach is not convex. This fact implies that the solver may converge to a local minimum, resulting in an overestimation of the bound, $\tilde{\beta}_d\geq \beta_d$. Such overestimation is prohibitive, as qubit systems could violate the associated dimensionality witness for the measurement settings of the global minimum. In order to certify its value, one thus needs to approach the bound $\beta_d$ from below. To this end, we will adapt a convex relaxation of the problem based on hierarchies of semidefinite programs (SDPs) \cite{bookSDP}, which are a class of well-behaved optimization problems efficiently solved by available routines \cite{Mosek}. \\


The second approach we propose consists in relaxing the minimization Eq.~\eqref{eq:probquditbound} with a sequence of SDPs. We outline here the method, additional details can be found in Supplementary Note 2.

The starting point consists of treating the measurement settings $\hat{\pi}$ as non-commuting symbols. 
Then, we assemble these and their products up to a certain order $l$ in a list, $\hat{\mathbf{m}} = \{1, \hat{\pi}, \hat{\pi}^2,..., \hat{\pi}^l \}$, where e.g. $\hat{\pi}^2$ contains all the degree 2 monomials, including $\hat{\pi}_{0|0}^2$, $ \hat{\pi}_{0|0}\hat{\pi}_{11}$, etc. 
This list of monomials serves as basis to construct the moment matrix $\Gamma = \langle \hat{\mathbf{m}}^\dagger \hat{\mathbf{m}}\rangle$, which constitutes the central object of this approach. 
In this way, each entry of $\Gamma$ corresponds to the expectation value of a monomial. 
Note that, by construction, the moment matrix is Hermitian, $\Gamma = \Gamma ^\dagger$, and positive semidefinite, $\Gamma \succeq 0$. The objective function $\langle p(\hat{\pi})\rangle$ can now be expressed as a linear combination of the entries of $\Gamma$, namely as $\langle p(\hat{\pi})\rangle = \mathrm{Tr}(C\Gamma)$ with $C$ a matrix of real coefficients. 
Similarly, the PSD constraints $\{\hat{\pi}_{a|x}\succeq 0 , \sum_{a}\hat{\pi}_{a|x} = \mathbb{1} \}$ are translated into linear matrix inequalities $\Lambda(\Gamma)\succeq 0$, with $\Lambda$ a corresponding (hermiticity-preserving) linear map. Thus, the problem we want to solve takes now the form   
\begin{equation}
    \begin{array}{crl} \min_{\Gamma\succeq 0}& \mathrm{Tr}(C\Gamma) \\
    \mbox{s.t.}& \Lambda(\Gamma)\succeq 0
    \end{array} \;,
    \label{eq:SDPpoly}
\end{equation}
which is a SDP.  

Note that, in the problem Eq.~\eqref{eq:SDPpoly} we have not yet introduced the parameter $d$, that is, the local Hilbert space dimension is still unspecified. Consequentially, the resulting value will lower bound $\beta_\infty$, and the inequality $\langle\hat{\mathcal{B}}\rangle\geq n\beta_\infty$ will be necessarily fulfilled for any data with a quantum explanation. This can be understood as a single-particle relaxation of the NPA algorithm introduced in Ref.~\cite{NavascuesNJP2008} to approximate the Tsirelson bound.   
We evaluate the bound for our example PIBI Eq.~\eqref{eq:THEinequality} and obtain $\beta =- 1/2$, which coincides with the qutrit bound presented in the previous paragraph. Such a bound can also be recovered analytically (see Supplementary Note 2). 

The next step involves introducing the constraints on the local Hilbert space dimension. For this purpose, several strategies are reported in the literature, such as the use of matrix polynomial identities or sampling methods \cite{navascues2015bounding}. {We adopt the latter approach, namely the Navascués–Vértesi hierarchy, implemented via the public package qdimsum, which is based on Ref.~\cite{TavakoliPRL2019}.}

First, we sample qudit POVMs for a specific $d$ and a state $\hat{\rho}$, compute the corresponding moment matrix $\Delta_1$. We do so until a basis of moment matrices $\{\Delta_i \}$ feasible by qudit data is completed. Then, the constraint $\Gamma = \sum_i f_i\Delta_i$, with $\{f_i\}$ free real coefficients, restricts $\Gamma$ onto the intersection of the PSD cone with the subspace of dimensionality constrained data. By applying such method to the PIBI Eq.~\eqref{eq:THEinequality} for qubits $d=2$, we recover the variational bound $\beta = -1/4$ at level $l=3$ of the SDP hierarchy. 

The relaxation proposed here gives a rigorous bound. However, it is potentially more memory demanding than the variational approaches { as it generally may require the consideration of high level $l$ (i.e., large moment matrices).} \\

\subsection*{Assessing the bounds}

We close this section by comparing the scalable bound for qubits and for qutrits with the Bell expectation value that can be realized exactly by the optimal settings returned from the variational optimization Eqs.~(\ref{eq:qubitoptproj},\ref{eq:qutritoptproj}). 

The Bell operator associated to PIBI Eq.~\eqref{eq:THEinequality} can be expressed as 
\begin{equation}
    \hat{\mathcal{B}} = \hat{T}_0 + \hat{T}_1^2 + \hat{T}_2^2,
    \label{eq:BellOpT}
\end{equation}
for $\{\hat{T}_a = \sum_{i\in[N]}\hat{t}_a^{(i)}\}$ certain collective operators [cf. Eq.~\eqref{eq:BellOp}]. If the optimal measurements are chosen, the single-particle operators for the qubit case read 
\begin{equation}
\begin{split}
    \hat{t}_0 &=  -(1/4)\ketbra{0}{0} + (3/4)\ketbra{1}{1} +2\ketbra{2}{2} \\
    \hat{t}_1 &= (\sqrt{3}/2)\ketbra{0}{1} + \mathrm{h.c.}  \\
    \hat{t}_2 &= 0    \ ,
\end{split}\ ,
\end{equation} 
and for qutrit
\begin{equation}
\begin{split}
    \hat{t}_0 &= -(1/2)\ketbra{0}{0} + \ketbra{1}{1} + (1/2)\ketbra{2}{2} \\
    \hat{t}_1 &= \ketbra{0}{1} + \mathrm{h.c.}     \\
    \hat{t}_2 &= (1/\sqrt{2})\ketbra{0}{2} + \mathrm{h.c.} 
\end{split} \ ,
\label{eq:qutrit_ts}
\end{equation} 
where, for simplicity, we have written them in the basis $\{\ket{0},\ket{1},\ket{2}\}$ where $\hat{t}_0$ is diagonal. 

Next, we represent the Bell operator Eq.~\eqref{eq:BellOpT} in the totally symmetric subspace of $n$ qutrits. Such subspace is of dimension $\binom{n+2}{2} = (n+2)(n+1)/2$ and it is spanned by the qutrit Dicke states \cite{AloyQMP}
\begin{equation}
   \ket{S_{\boldsymbol{\mu}}} ={n\choose \boldsymbol{\mu}}^{-1/2} \sum_{\pi \in \mathfrak{S}_{n}} \pi\left(\ket{0}^{\otimes \mu_0}(\ket{1}^{\otimes \mu_1}(\ket{2}^{\otimes \mu_2}\right) ,
\label{eq:Dicke_basis}
\end{equation}
where $\boldsymbol{\mu} = (\mu_0, \mu_1, \mu_2)$ is an partition of $n$, i.e. a set of integers adding up to $n$, $\mu_0 + \mu_1 + \mu_2 = n$, ${n\choose \boldsymbol{\mu}} = n!/(\mu_0!\mu_1!\mu_2!)$ is the multinomial coefficient, and the sum is over distinct permutations $\pi\in\mathfrak{S}_{n}$. In physical terms, $\{\mu_0, \mu_1, \mu_2 \}$ represent the occupation number onto levels $\{\ket{0},\ket{1}, \ket{2} \}$, respectively. Since the total number of particle is fixed, we take $\mu_0$, $\mu_1$ to be the independent variables describing the state and set $\mu_2 = n-\mu_0-\mu_1$.

Then we probe the minimal eigenvalue of the Bell operator, $\lambda_{\min}(\hat{\mathcal{B}})$, for different number of parties $n$. If $\hat{\mathcal{B}}$ is interpreted as a Hamiltonian \cite{FadelQuantum2018}, $\lambda_{\min}(\hat{\mathcal{B}})$ is its ground state energy. 
The results are shown in Figure~\ref{fig:boundsn} (blue dots for qubit and orange stars for qutrit). In the same Figure, we also display an analytical approximation of the previous minimal eigenvalues (solid blue line for qubits and orange dashed line for qutrits), which become tight in the thermodynamic limit $n\rightarrow \infty$. These are based on the Holstein-Primakoff (HP) mapping, which we outline below. \\

From the structure of the Bell operator Eq.~\eqref{eq:BellOpT}, in order to minimize its expectation value one has to minimize the mean value of $\hat{T}_0$ while minimizing the fluctuations (second moments) of $\hat{T}_1$ and $\hat{T}_2$ as well. { Thus, it corresponds to a squeezed state along components $\hat{T_1},\hat{T_2}$. Later, we exploit this fact to relate the violation of the witness to metrologically useful states as prepared in spin-1 BECs.} Notice that the ground state of $\hat{T}_0$ corresponds to all parties occupying the same state, $\ket{0}$, i.e. $\ket{0}^{\otimes n} = \ket{S_{[n,0,0]}}$ as ``the winner takes it all''. For large system sizes, the effect of $\hat{T}_1$, $\hat{T}_2$ can be treated perturbatively. To this end, we consider $\ket{S_{[n,0,0]}}$ as the vacuum of a two-mode oscillator with creation operators $a_1^\dagger$, $a_2^\dagger$, exciting parties from $\ket{0}$ to $\ket{1}$ and $\ket{2}$, respectively \cite{FadelQuantum2018}. With this ansatz, one can linearize $\hat{\mathcal{B}}$ and obtain an analytical expression for the ground energy (see Supplementary Note 3)
\begin{equation}
\begin{split}
E_2 & =   -\frac{n}{4}+\frac{1}{2}(\sqrt{3n+1}-1) \\ 
E_3& =  -\frac{n}{2}- \frac{5}{4}+\sqrt{\frac{3n}{2} + \left(\frac{3}{4}\right)^2} + \sqrt{\frac{n}{2} + \left(\frac{1}{2}\right)^2}   \;.
\label{eq:Bogobounds_main}
\end{split}
\end{equation}
In such a limit, $n\gg 1$, the optimal settings for $\langle\hat{\beta}\rangle$ converge to the optimal settings for the full Bell operator. As a consequence, Eqs.~\eqref{eq:Bogobounds_main} are to be understood as finite size corrections to the scalable bounds addressed in the previous subsection. As a byproduct of this calculation, we verify that in the thermodynamic limit the (tight) quantum bound of PIBI Eq.~\eqref{eq:THEinequality} is indeed $-n/2$, and that it can be saturated by qutrits. 
The measurements considered here lose their optimality for small system sizes. Such finite-size effect can be appreciated in Figure~\ref{fig:boundsn}, where notice qubit states with stronger violation than qutrit.

\begin{figure}[h]
\centering
\includegraphics[width=0.99\linewidth]{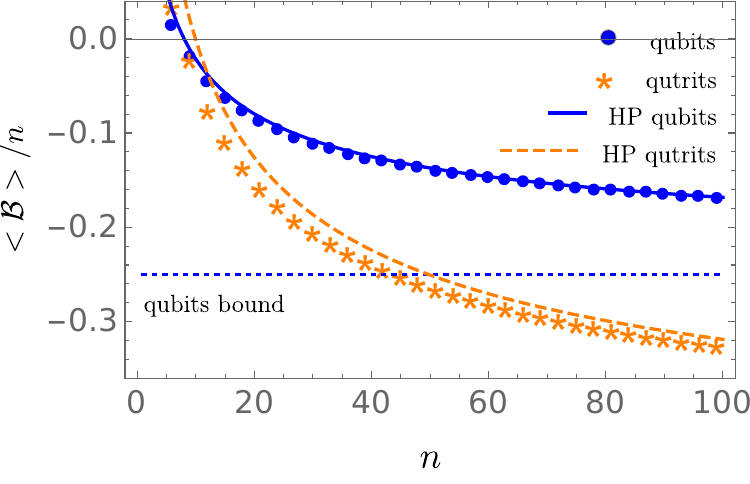}
\caption{
Bounds on the Bell expectation value as a function of the number of parties $n$. Any expectation value below the classical bound (zero, thin black line) imply Bell correlations. The blue dots represent the minimal Bell expectation value reachable by performing on totally symmetric states of qubit the optimal settings specified in Eq.~\eqref{eq:qubitoptproj}. The orange stars are the minimal Bell expectation value from measurements  Eq.~\eqref{eq:qutritoptproj} performed on totally symmetric states of qutrit. Finally, the dotted horizontal line (blue) indicates the scalable qubit bound derived in  Eq.~\eqref{eq:DimBell}. }
\label{fig:boundsn}
\end{figure}

From Figure~\ref{fig:boundsn}, we observe how the qutrit states surpass the qubit bound when $n \gtrsim 54$, which signals the detection of correlations that cannot result from qubit states. From Eq.~\eqref{eq:n_d}, we establish the number of high dimensional systems (qutrits) that are certified from the violation achievable by two-mode bosonic states $E_3$ (dashed orange line of Figure~\ref{fig:boundsn}).  Given that $\beta_2 = -1/4, \beta_\infty = \beta_3 = -1/2$ (as discussed in the previous subsection) and $\langle \hat{\mathcal{B}}\rangle = E_3$, the number of certified qutrits reaches  
\begin{equation}
    n' = \lceil n - 4(E_3+n/2) \rceil
   \label{eq:nqutrits}
\end{equation}
It is important to stress at this point that the two-mode bosonic state is used to generate the necessary correlations to achieve the violation of the PIBI presented in Figure~\ref{fig:boundsn}. The number of certified qutrits (Eq.~\eqref{eq:nqutrits}) is independent of any such detailed assumptions about systems and only relies on the statistics enabling the PIBI violation. In the many-body regime, $n\rightarrow\infty$, where the Hilbert space becomes unbounded, one might naively expect the single-particle dimension to be irrelevant. That is, the accessible correlations by qubits and qutrits converging to the same set for sufficiently large $n$. However, the expectational robustness of our PIBI enables the certification of a ``qutrit fraction'' $n'/n \sim 1 -\kappa$, with $\kappa = (E_3 + n/2)/n$. Note that $\kappa\geq 0$ is of the order $\mathcal{O}(n^{-1/2})$. Thus, in the thermodynamic limit $n\rightarrow\infty$, remarkably, all parties must be qutrits. In addition, we note that in such a limit, the violation of the Bell dimension witness is extensive in the system's size. This observation underscores the advantages of implementing the criterion experimentally, as opposed to other e.g. bipartite witnesses is which the gap between qubit and qutrit correlations is narrower~\cite{Gallego2010}. Finally, while it is known in condensed matter physics that extra individual degrees of freedom (e.g., spin) give rise to qualitatively different phases in the macroscopic scale, here we unveil a similar result from a device independent perspective by confirming the fact that certain correlations in many-body systems require high-dimensional quantum resources to be explained.

Having established the bounds, we now move to characterize further the states that maximally violate PIBI Eq.~\eqref{eq:THEinequality}, or its associated Bell dimension witness Eq.~\eqref{eq:DimBell}. \\

{In order to gain insight about the state $\ket{\Psi}$ that maximally violates the Bell inequality, we exploit the freedom to apply local unitaries without loss of generality. We found a unitary $\hat{U}$ (see Supplementary Note 4), which makes the equivalent state $\ket{\Psi'} = \hat{U}^{\otimes n}\ket{\Psi}$ specially illustrative. We represent such a state in Figure~\ref{fig:illuminati} by plotting the absolute value of its coefficients with respect to the Dicke basis Eq.~\eqref{eq:Dicke_basis}. } 

\begin{figure}[h]
\centering
\includegraphics[width=0.9\linewidth]{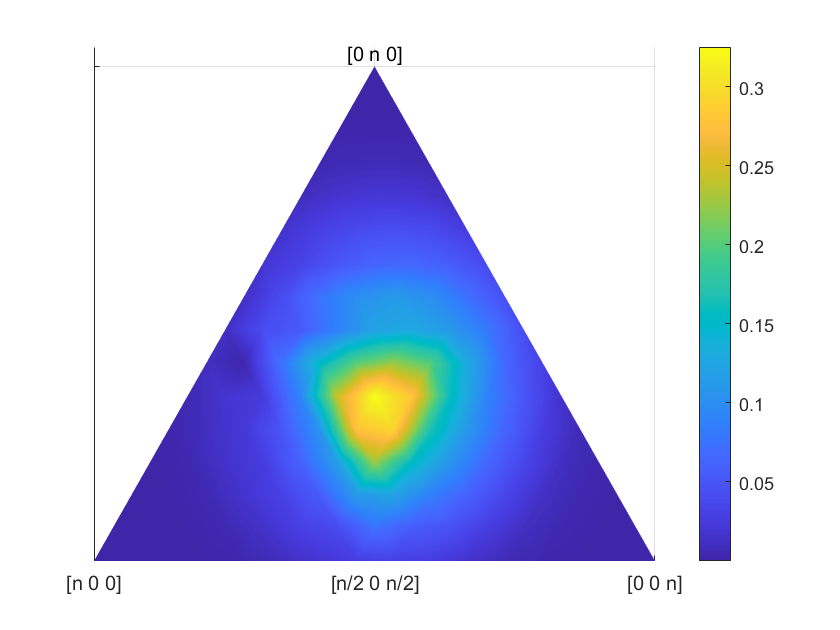}
\caption{Visualization of a state $\ket{\Psi'}$ maximally violating the Bell inequality with the qutrit settings specified in the text and  $n = 15$ parties. The color scale indicates the overlap with the Dicke basis $|\braket{\Psi'}{S_{\boldsymbol{\mu}}}|$. { The vertices of the triangle correspond to the overlap with the Dicke states $\{\ket{S_{[n,0,0]}} = \ket{0}^{\otimes n},\ket{S_{[0,n,0]}} = \ket{1}^{\otimes n},\ket{S_{[0,0,n]}} = \ket{2}^{\otimes n}\}$, while the distribution is peaked around the center, which corresponds to the balanced Dicke state, $\ket{S_{[n/3,n/3,n/3]}}$.    }}
\label{fig:illuminati}
\end{figure}

From Figure~\ref{fig:illuminati}, we see how the state $\ket{\Psi'}$ has non-vanishing support in all three levels and how it is almost invariant under their permutations. Such observations support the fact that it is indeed genuinely three-level correlated. Moreover, the coefficients follow an approximately Gaussian distribution with mean at the balanced Dicke state. Similar Gaussian states can be found in spin-1/2 Lipkin models for low temperatures~\cite{Marconi2022}, or in ground states of spin-1 condensates~\cite{You2002}. 

\subsection*{Derivation of Eq.~\eqref{eq:spin_nematic_witness}}
We proceed to derive the witness Eq.~\eqref{eq:spin_nematic_witness}.
With the measurement settings defined in Eq.~\eqref{eq:settings2}
\begin{align}
   \label{eq:Bthetqubits}
    \hat{\mathcal{B}}(\theta) &= \cos^2(\theta)\hat{\Lambda}_0^2 + \sin(\theta)\hat{\Lambda}_1 + \sin^2(\theta)(n-\hat{n}_2) + 2\hat{n}_2  \;,
\end{align}
which is written in terms of the collective Gell-Mann matrices $\Lambda_a = \sum_{i\in [n]}\hat{\lambda}_a(i)$ and of the population onto level $\ket{2}$, $\hat{n}_2 = \sum_{i\in [n]}\ketbra{2}(i)$.

In terms of the shorthand notation introduced in the main text, the expectation value of Eq. \eqref{eq:Bthetqubits} is 
\begin{equation}
   \frac{\langle\hat{\mathcal{B}}(\theta)\rangle}{nz}  = \cos^2(\theta)x + \sin(\theta)y + \sin^2(\theta) + \frac{2(1-z)}{z} \ .
    \label{eq:BellOpxyz}
\end{equation}
For $y^2/30\leq x\leq (2-y)/2$, the optimal angle to minimize Eq.~\eqref{eq:BellOpxyz} 
fulfills
\begin{equation}
\sin(\theta) = y/[2(x-1)] \ .
\label{eq:thetaopt}
\end{equation}
From Eq.~\eqref{eq:BellOpxyz}, we obtain the inequality
\begin{equation}
    2-\beta + (2x-3)z + \sqrt{(\beta + z - y z-2) (\beta + z + y z-2)} \geq 0 \ ,
\end{equation}
where $\beta\leq \langle\hat{\mathcal{B}} \rangle/n $. For $\beta = 0$, we obtain the Bell correlation witness (multiplied by $z$). Similarly, for $\beta = -1/4$, one would obtain the dimension witness 
\begin{equation}
    9 + 4z(2 x-3)  + \sqrt{9^2 - 8 z (9 + 2 z(y^2-1))}\geq 0 \;.
    \label{eq:dimwittrivial}
\end{equation}
However, since the maximal violation of the Bell operator considered here is achieved by qubits, the dimension witness Eq.~\eqref{eq:dimwittrivial} is trivial. Indeed, at the point of maximal violation $z= 1$, which results in $8x+\sqrt{5^2 - (4y)^2}\geq 3$. For a perfectly squeezed state, $x = 0$ and $y = 1$, then, $\sqrt{5^2-4^2} = 3$ is saturated. \\

\subsection*{Derivation of witness Eq.~(22)}
\label{app:su3squeezing}

Here we proceed to derive Eq.~(22) of the Results section. \\

We consider the following convention to label outcomes of the individual observable $\hat{m}_x$
\begin{equation}
\begin{split}
    \hat{m}_x :=& \hat{\pi}(0|x) - \hat{\pi}(1|x)  \\
    \hat{q}_{xx} =&  \hat{\pi}(0|x) + \hat{\pi}(1|x) = \{\hat{m}_x,\hat{m}_x\}/2 \ .
\end{split}
\end{equation}
Inverting the above relations, we obtain
\begin{equation}
\begin{split}
    \hat{\pi}(0|x) =& (\hat{q}_{xx}+\hat{m}_x)/2  \\
    \hat{\pi}(1|x) =& (\hat{q}_{xx} -\hat{m}_{x})/2  \;.
\end{split}
\end{equation}

From here, one can rewrite the Bell operator from moments of collective observables $\hat{M}_x = \sum_{i\in[n]}\hat{m}_{x}{(i)}$ (and collective moments of local observables, $\hat{Q}_{xy} = \sum_{i\in[n]}\hat{q}_{xy}{(i)}$, $\hat{\mathfrak{Q}}_{xy} = (1/2)\sum_{i\in[n]}\hat{q}_{xx}\hat{q}_{yy}{(i)}+\mathrm{h.c.} $). 
For the one and two-body terms we have  
\begin{equation}
\begin{split}
\hat{\Pi}_{01|01}+\hat{\Pi}_{01|10} &=\hat{\mathfrak{Q}}_{01}-\hat{Q}_{01} \\
 2[(\hat{\Pi}_{0|0}-\hat{\Pi}_{1|1})^2+(\hat{\Pi}_{0|1}-\hat{\Pi}_{1|0})^2] & =  (\hat{M}_0 + \hat{M}_1)^2 \\
 &+ (\hat{Q}_{00}-\hat{Q}_{11})^2 \ ,
\end{split}
\end{equation}
yielding 
\begin{equation}
    \hat{\mathcal{B}} = [(\hat{M}_0 + \hat{M}_1)^2 + (\hat{Q}_{00}-\hat{Q}_{11})^2]/2 + \hat{\mathfrak{Q}}_{01}-\hat{Q}_{01} \;.
\label{eq:BellOpMQ}
\end{equation}

Next, we exploit local spin-1 measurements along direction specified by $\mathbf{r}\in \mathbb{R}^3$, namely $\hat{m} = \mathbf{r}\cdot\hat{\mathbf{s}}$. As explained in the main text, we consider two measurement settings with angle $2\theta$ between them
\begin{equation}
\begin{split}
\hat{m}_0 &= \cos(\theta)\hat{s}_x + \sin(\theta)\hat{s}_y \\
\hat{m}_1 &= \cos(\theta)\hat{s}_x - \sin(\theta)\hat{s}_y 
\end{split}
\end{equation}
where $\hat{s}_x$ and $\hat{s}_y)$ are spin-1 operators (or any SU(3) equivalent operators) along two orthogonal directions.

From these, we derive 
\begin{equation}
\begin{split}
\hat{q}_{00} &=   c^2\hat{q}_{xx} +  s^2\hat{q}_{yy} +2cs\hat{q}_{xy}    \\
 \hat{q}_{11} &=  c^2\hat{q}_{xx} +  s^2\hat{q}_{yy} -2cs \hat{q}_{xy}       \\
 \hat{q}_{01} &= c^2\hat{q}_{xx} - s^2\hat{q}_{yy}  \ .
\end{split}
\end{equation}
where we have used the shorthand notation $c=\cos(\theta)$ and $s = \sin(\theta)$. These expressions are inserted into Eq.~\eqref{eq:BellOpMQ}, and the properties of $\mathfrak{su}(3)$ matrices are used to linearize the anticommutators appearing in e.g. $\hat{\mathfrak{Q}}$. As a result, the Bell operator reads
\begin{equation}
\begin{split}
    \hat{\mathcal{B}}(\theta) &= 2c^2\hat{S}_x^2  \\  
   & - s^2c^2(3\hat{Q}_{zz} + \hat{Q}_{xx} -8\hat{Q}_{xy}^2-2n\mathbb{I}) \\
    & + s^2(1+s^2)\hat{Q}_{yy} \ .
\end{split}
\end{equation}
The optimal angle $\theta^*$ may be calibrated from the data $\langle\hat{S}_x^2 \rangle:=xn$, $\langle 3\hat{Q}_{zz} + \hat{Q}_{xx} -8\hat{Q}_{xy}^2-2n\mathbb{I}\rangle:= yn$, $\langle \hat{Q}_{yy} \rangle:= zn$ and is set from $s^2 = (2x+y-z)/[2(y+z)]$ in order to derive Eq.~(22) from $\langle \hat{\mathcal{B}}(\theta = \theta^*) \rangle\geq n\beta$. \\

 Here, we will address the optimization of witness Eq.~(22) to detect Bell correlation and qutrit correlations in the time-propagated state $\ket{\Psi} = e^{-it\hat{H}}\ket{0}^{\otimes n}$ with the spin-1 BEC Hamiltonian. To this end, we will construct a single-party unitary $U$ which applied to the state as per $U^{\otimes n}\ket{\Psi}$ will bring $\langle \hat{S}_x^2 \rangle, \langle \hat{Q}_{xy}^2 \rangle$ to the two squeezed second moments in the $\{\hat{S}_x, 2\hat{Q}_{yz}^2, \hat{Q}_{zz}-\hat{Q}_{yy} \}$, $\{\hat{S}_y, 2\hat{Q}_{xz}, \hat{Q}_{xx}-\hat{Q}_{zz}\}$ subspaces.  

Such an optimal unitary reads: 
\begin{equation}
    U = \begin{pmatrix}
    -(v_{xz}+iv_{y})/2 & 1/\sqrt{2} &   (v_{xz}+iv_{y})/2 \\
    (v_x - iv_{yz})/\sqrt{2} & 0 &  (v_x - iv_{yz})/\sqrt{2}  \\
    (v_{xz}+iv_{y})/2 & 1/\sqrt{2} &    -(v_{xz}+iv_{y})/2
    \end{pmatrix}
\end{equation}
where $i$ is the imaginary unit and $(v_x, v_{yz})$ is the (normalized) minimal eigenvector of 
\begin{equation}
   C = 
\begin{pmatrix}
    \langle \hat{S}_x^2 \rangle &     2\Re{\langle \hat{S}_x\hat{Q}_{yz} \rangle} \\
     2\Re{\langle \hat{S}_x \hat{Q}_{yz} \rangle} &  4\langle \hat{Q}_{yz}^2 \rangle 
\end{pmatrix} \;,
\end{equation}
and $(v_y, v_{xz})$ is the (normalized) minimal eigenvector of 
\begin{equation}
   C' = 
\begin{pmatrix}
    \langle \hat{S}_y^2 \rangle &     2\Re{\langle \hat{S}_x\hat{Q}_{xz} \rangle} \\
     2\Re{\langle \hat{S}_x \hat{Q}_{xz} \rangle} &  4\langle \hat{Q}_{xz}^2 \rangle 
\end{pmatrix} \ .
\end{equation}

The state $\ket{\Psi}$ remains as eigenstate of $\hat{S}_z$ during time evolution. Thus, the squeezing in the two subspaces is identical, and, in particular, $v_x = v_y, v_{yz} = -v_{xz}$. However, our witness can support different squeezing strengths as well, resulting from experimental imperfections or a different initial state.

\section*{DATA AVAILABILITY}

 The data to construct the plots are available at the following link \url{https://github.com/GuillemMRR/3-PIBIs_dimW}. \\

\section*{CODE AVAILABILITY}

 The code is available at the following link \url{https://github.com/GuillemMRR/3-PIBIs_dimW}. \\

\section*{ACKNOWLEDGMENTS}

We thank Irénée Frérot for useful discussions. 
GM and ML acknowledge support from: Europea Research Council AdG NOQIA; MCIN/AEI (PGC2018-0910.13039/501100011033, CEX2019-000910-S/10.13039/501100011033, Plan National FIDEUA PID2019-106901GB-I00, Plan National STAMEENA PID2022-139099NB, I00, project funded by MCIN/AEI/10.13039/501100011033 and by the “European Union NextGenerationEU/PRTR" (PRTR-C17.I1), FPI); QUANTERA MAQS PCI2019-111828-2); QUANTERA DYNAMITE PCI2022-132919, QuantERA II Programme co-funded by European Union’s Horizon 2020 program under Grant Agreement No 101017733); Ministry for Digital Transformation and of Civil Service of the Spanish Government through the QUANTUM ENIA project call - Quantum Spain project, and by the European Union through the Recovery, Transformation and Resilience Plan - NextGenerationEU within the framework of the Digital Spain 2026 Agenda; Fundació Cellex; Fundació Mir-Puig; Generalitat de Catalunya (European Social Fund FEDER and CERCA program, AGAUR Grant No. 2021 SGR 01452, QuantumCAT \ U16-011424, co-funded by ERDF Operational Program of Catalonia 2014-2020); Barcelona Supercomputing Center MareNostrum (FI-2023-3-0024); Funded by the European Union. Views and opinions expressed are however those of the author(s) only and do not necessarily reflect those of the European Union, European Commission, European Climate, Infrastructure and Environment Executive Agency (CINEA), or any other granting authority. Neither the European Union nor any granting authority can be held responsible for them (HORIZON-CL4-2022-QUANTUM-02-SGA PASQuanS2.1, 101113690, EU Horizon 2020 FET-OPEN OPTOlogic, Grant No 899794), EU Horizon Europe Program (This project has received funding from the European Union’s Horizon Europe research and innovation program under grant agreement No 101080086 NeQSTGrant Agreement 101080086 — NeQST); ICFO Internal “QuantumGaudi” project; European Union’s Horizon 2020 program under the Marie Sklodowska-Curie grant agreement No 847648; “La Caixa” Junior Leaders fellowships, La Caixa” Foundation (ID 100010434): CF/BQ/PR23/11980043.

AA acknowledges support from the Austrian Science Fund (FWF) (projects P 33730-N and 10.55776/PAT2839723) and by the ESQ Discovery programme (Erwin Schr{\"o}dinger Center for Quantum Science \& Technology), hosted by the Austrian Academy of Sciences ({\"O}AW).

MF was supported by the Swiss National Science Foundation Ambizione Grant No. 208886, and The Branco Weiss Fellowship -- Society in Science, administered by the ETH Z\"{u}rich.

JT has received support from the European Union’s Horizon Europe program through the ERC StG FINE-TEA-SQUAD (Grant No. 101040729). JT also acknowledges support from the Quantum Delta NL program. This publication is part of the ‘Quantum Inspire – the Dutch Quantum Computer in the Cloud’ project (with project number [NWA.1292.19.194]) of the NWA research program ‘Research on Routes by Consortia (ORC)’, which is funded by the Netherlands Organization for Scientific Research (NWO). \\

\section*{AUTHOR CONTRIBUTIONS}
AA, JT, ML obtained the Bell Inequality introduced in this work and studied its violations (Figure~5).
GM made the remaining contributions and wrote the manuscript, assisted by MF. All authors read and approved the manuscript. \\

\section*{COMPETING INTERESTS}
The authors declare no competing financial or non-financial interests. \\

\section*{REFERENCES}
\bibliographystyle{naturemag} 
\bibliography{BellDim.bib}

\end{document}